\PassOptionsToPackage{dvipsnames,svgnames,table}{xcolor}

\documentclass[a4paper,11pt]{article}

\usepackage{jheppub}
\usepackage{hyperref}
\usepackage{orcidlink}

\usepackage{lineno}
\usepackage{graphicx}
\usepackage{mathtools}
\usepackage{braket}
\usepackage{pifont}

\usepackage{tikz}
\usepackage[compat=1.1.0]{tikz-feynman}
\usepackage{contour}
\usepackage{enumitem}

\newcommand{\alp}{\alpha^\prime}
\newcommand{\pa}{\partial}
\newcommand{\Kin}{{\mathrm K}}
\newcommand{\be}{\begin{equation}}
\newcommand{\ee}{\end{equation}}

\title{High-Energy Fixed-Angle Meson Scattering and\\the Constituent Counting Rule in Holographic QCD}

\author[a]{Adi Armoni,}
\affiliation[a]{Department of Physics, Swansea University,\\Singleton Park, Swansea SA2 8PP, UK}
\emailAdd{a.armoni@swansea.ac.uk}

\author[b, c]{Bartosz Pyszkowski\,\orcidlink{0009-0004-3144-1842},}
\affiliation[b]{Department of Physics, Kyoto University,\\Kitashirakawa Oiwakecho, Sakyo-ku, Kyoto 606-8502, Japan}
\affiliation[c]{Center for Gravitational Physics and Quantum Information,\\Yukawa Institute for Theoretical Physics, Kyoto University,\\Kitashirakawa Oiwakecho, Sakyo-ku, Kyoto 606-8502, Japan}
\emailAdd{pyszkowski.bartosz@yukawa.kyoto-u.ac.jp}

\author[b, c, d]{Shigeki Sugimoto,}
\affiliation[d]{Kavli IPMU (WPI), UTIAS, The University of Tokyo,\\Kashiwa, Chiba 277-8582, Japan}
\emailAdd{sugimoto@gauge.scphys.kyoto-u.ac.jp}

\author[e]{and Dorin Weissman\,\orcidlink{0000-0001-9697-0252}}
\affiliation[e]{Asia Pacific Center for Theoretical Physics,\\Cheongam-ro, Nam-gu, Pohang 37673, Republic of Korea}
\emailAdd{dorin.weissman@apctp.org}

\abstract{We investigate the high-energy fixed-angle scattering of pions and $\rho$-mesons~in a~bottom-up holographic QCD model.\,\,To this end, we generalise the approach of Polchinski and Strassler \cite{Polchinski:2001tt} to write an ansatz for meson scattering amplitudes based on superstring scattering amplitudes in asymptotically AdS space.\,\,We demonstrate that our generalisation of the Polchinski--Strassler proposal is necessary to describe $\rho$-meson scattering consistently with the Nambu--Goldstone boson equivalence theorem.\,\,Our results for pion and $\rho$-meson scattering amplitudes are in agreement with the constituent counting rule found in QCD. Moreover, our proposal for 2-to-2 scattering amplitudes provides a method for computing scattering angle dependence.}

\begin{document}
\begin{flushright}
    KUNS-3032\\
    YITP-24-181
\end{flushright}

\maketitle
\flushbottom
\clearpage
\section{Introduction} \label{sec:1_Intro}

The early days of string theory can be traced back to 1968 when Veneziano first wrote down a cross-symmetric, Regge-behaved scattering amplitude meant to describe 2-to-2 scattering of low-lying mesons \cite{Veneziano:1968yb} (with some of the clues to the result found earlier in reference \cite{Ademollo:1968cno}). The cross-symmetric property of strong interactions, also known at the time as the duality hypothesis or the Dolen, Horn, and Schmid duality, suggested that the $s$- and $t$-channels of meson scattering amplitudes should be equivalent \cite{Dolen:1967zz, Dolen:1967jr, Schmid:1968zz, Green:1987sp}.\,\,Moreover, in the Regge regime, where $s$ is large and $t$ is fixed, the scattering amplitudes were required to be $\mathcal{A} \sim s^{\alpha(t)}$. Both requirements were captured by the Veneziano formula:
\begin{equation} \label{eq:1_V_Formula}
    \mathcal{A}(s, t) = \frac{\Gamma(-\alpha(s)) \, \Gamma(-\alpha(t))}{\Gamma(-\alpha(s)-\alpha(t))},
\end{equation}
where $\alpha(s)$ and $\alpha(t)$ denote linear functions of $s$ and $t$.

Around the time Veneziano published his formula, experiments conducted at the SLAC National Accelerator Laboratory investigating proton- and neutron-electron scattering at large $s$ and $t$ with their ratio fixed (i.e.\,\,in the high-energy fixed-angle regime) found that the differential cross sections for these processes exhibit power-law scaling with $s$ \cite{Coward:1967au, PhysRevLett.23.930, PhysRevLett.23.935, Litt:1969my, Bodek:1973dy, Poucher:1973rg}. Famously, these results reinforced the idea that hadrons are composite particles made up of point-like constituents collectively referred to as partons \cite{Feynman:1969ej, Bjorken:1969ja}.

On the other hand, the Veneziano formula \eqref{eq:1_V_Formula} in the high-energy fixed-angle regime displays exponentially soft behaviour in $s$, and although the above-mentioned experiments at SLAC focused on proton- and neutron-electron scattering, it was readily recognised that elastic, high-energy fixed-angle meson scattering would also show power-law scaling with $s$. This, in consequence, invalidated the application of the Veneziano formula to the scattering of low-lying mesons in the high-energy fixed-angle regime.

Soon after the SLAC experiments, dimensional counting arguments in four-dimensional asymptotically free confining theories (e.g.\,\,in four-dimensional QCD) led to a prediction for the leading-order scaling with $s$ of exclusive hadronic amplitudes in the high-energy fixed-angle regime, which matched the expected behaviour \cite{Matveev:1973ra, Brodsky:1973kr, Brodsky:1974vy}:
\begin{equation} \label{eq:1_Scaling_Rule}
    \mathcal{A}(s, \theta_i) \sim s^{2-\frac{m}{2}} f(\theta_i),
\end{equation}
where $m$ is the minimum number of hard constituents involved in the scattering process, and $f(\theta_i)$ captures the dependence of $\mathcal{A}$ on the scattering angles $\theta_i$.\,\,Equation \eqref{eq:1_Scaling_Rule} is known in the literature as the constituent counting rule.

Today, with the aid of holography \cite{Maldacena:1997re}, we can reconsider whether string scattering can give rise to power-law scaling with $s$ in the high-energy fixed-angle regime.\,\,An early breakthrough in this direction came from Polchinski and Strassler \cite{Polchinski:2001tt}, who put forward a~proposal based on a Virasoro--Shapiro-like amplitude formulated in hard-wall AdS space that appears to recover the leading-order scaling with $s$ for $n$-point scalar glueball scattering amplitudes in agreement with the constituent counting rule.\,\,The problem of recovering the~constituent counting rule and other partonic features from string scattering amplitudes was also explored in the context of holography in references \cite{Andreev:2002aw, Brower:2002er, Polchinski:2002jw, Andreev:2004sy, Hatta:2007he, Bianchi:2021sug}.

In reference \cite{Polchinski:2001tt}, Polchinski and Strassler further mentioned that their proposal can be used to study the scattering of states with arbitrary spin, in agreement with the constituent counting rule \eqref{eq:1_Scaling_Rule}.\,\,However, in this paper, based on a study of vector meson scattering in the context of a holographic QCD model in an asymptotically AdS space (to be introduced in \hyperref[sec:2_HQCD_Model]{Section 2}), we will show that naively using the proposal \cite{Polchinski:2001tt} fails to recover the constituent counting rule and, more importantly, disagrees with the Nambu--Goldstone (NG) boson equivalence theorem.\,\,To address this, we will introduce a new generalised proposal that is consistent with both the constituent counting rule and the NG boson equivalence theorem.

Although the precise details of our generalised proposal are not essential at this stage, briefly examining it can help clarify our objectives.\,\,In the context of our holographic QCD model from \hyperref[sec:2_HQCD_Model]{Section 2}, our proposal relates the high-energy fixed-angle limit of an $n$-meson scattering amplitude $\mathcal{A}_n$ in four dimensions to an $n$-superstring scattering amplitude~in a~five-dimensional asymptotically AdS (AAdS) space:
\begin{equation} \label{eq:1_PS_Gen}
    \mathcal{A}_n \Big(k_{\mu}^{(i)}, \zeta_{\mu}^{(i)}\Big) = \int dw \sqrt{-g} \times \widetilde{\mathcal S}_n \Big(p_M^{(i)}, \zeta_M^{(i)}\Big) \times \prod_{i=1}^n \psi^{(i)}(w),
\end{equation}
where $k_{\mu}^{(i)}$ and $\zeta_{\mu}^{(i)}$ for $i = 1, \ldots, n$ and $\mu = 0, \ldots, 3$ are the momenta and polarisations of the scattered mesons, respectively.\,\,The coordinates $x^M = (x^{\mu}, w)$ for $M = 0, \ldots, 4$ parameterise the AAdS space, $g$ is the AAdS metric determinant, and $\widetilde{\mathcal S}_n$ is a differential operator obtained from an $n$-superstring scattering amplitude $\mathcal{S}_n$ of gauge bosons in five-dimensional Minkowski space, acting on the wavefunctions $\psi^{(i)}(w)$ of the scattered mesons. Inside $\widetilde{\mathcal S}_n$, dot products are performed using the AAdS metric, and the fifth component~of the polarisations $\zeta_M^{(i)}$, i.e.\,\,$\zeta_w^{(i)}$, is assigned based on the type of meson.\,\,The momenta $p_M^{(i)}$ will be defined later, along with other relevant details.\,\,In addition, let us note that, compared to the original proposal for scalar glueballs \cite{Polchinski:2001tt}, the novel feature in our proposal \eqref{eq:1_PS_Gen} are the components $p_w^{(i)}$, which will be necessary to describe vector meson scattering.

In this paper, we will always work in the regime where $\alp/R_5^2 \ll 1$, with $\alp$ and $R_5$ denoting the Regge slope and radius of curvature of the five-dimensional AAdS space in its asymptotically AdS region, respectively.

The structure of this paper is as follows.\,\,In \hyperref[sec:2_HQCD_Model]{Section 2}, we outline the essential features of our bottom-up holographic QCD model.\,\,The main findings of this paper can be found in \hyperref[sec:3_PS_Prop]{Section 3} and \hyperref[sec:4_Scaling_Laws]{Section 4}.\,\,In \hyperref[sec:3_1_PS_Original]{Section 3.1}, we review the original proposal.\,\,In \hyperref[sec:3_2_PS_General]{Section 3.2}, we present our generalised proposal.\,\,Then, in \hyperref[sec:3_3_PS_GenEx]{Section 3.3}, we examine a toy model related to the holographic QCD model from \hyperref[sec:2_HQCD_Model]{Section 2} to motivate our generalised proposal.

In \hyperref[sec:4_Scaling_Laws]{Section 4}, we use both the original and generalised proposals in our holographic QCD model to determine the scaling with $s$ of pion and $\rho$-meson scattering amplitudes in the high-energy fixed-angle limit.\,\,In \hyperref[sec:4_1_Naive_Counting]{Section 4.1}, we illustrate how a naive application of the original proposal recovers the constituent counting rule for these scattering processes. In~\hyperref[sec:4_2_PS_Vectors]{Section 4.2}, we refine the previous analysis, finding that naively using the original proposal fails to recover the constituent counting rule for processes involving $\rho$-mesons, which, as we show, implies a contradiction with the NG boson equivalence theorem.\,\,In \hyperref[sec:4_3_Scaling_Corrected]{Section~4.3}, we demonstrate how our generalised proposal recovers the constituent counting rule for pion and $\rho$-meson scattering amplitudes, consistently with the NG boson equivalence theorem.

In \hyperref[sec:5_Conclusions]{Section 5}, we outline our findings and directions for future work.\,\,Four appendices are also included.\,\,In \hyperref[app:A_Scaling_Rules]{Appendix A}, we review the derivation of the constituent counting rule based on dimensional counting arguments.\,\,In \hyperref[app:B_NG_Theorem]{Appendix B}, we review the NG boson equivalence theorem.\,\,In \hyperref[app:C_pw_zero]{Appendix C}, we justify a statement from \hyperref[sec:3_2_PS_General]{Section 3.2}.\,\,In \hyperref[app:D_zeta_choice]{Appendix~D}, we briefly review some necessary details on polarisation vectors of massive vector bosons.

\subsection*{Conventions and Notation}

We use natural units, where $c = \hbar = 1$, and adopt the mostly-plus sign convention for the metric tensor.\,\,The metric tensor $g_{MN}$ is reserved for curved spaces, while $\eta_{\mu\nu}$ denotes the Minkowski metric.\,\,Uppercase Latin letters label Lorentz indices ranging from 0 to 4, whereas lowercase Greek letters represent Lorentz indices ranging from 0 to 3.\,\,We adopt the convention where all external momenta in a scattering process are treated as incoming. For $2$-to-$2$ scattering with initial momenta $k^{(1)}$ and $k^{(2)}$ and final momenta $k^{(3)}$ and $k^{(4)}$, we define the Mandelstam variables as follows:
\begin{equation}
    s = -\Big(k^{(1)}+k^{(2)}\Big)^2, \qquad t = -\Big(k^{(2)}+k^{(3)}\Big)^2, \qquad u = -\Big(k^{(1)}+k^{(3)}\Big)^2,
\end{equation}
in agreement with the conventions used in reference \cite{Green:1987sp}.
\section{A Holographic QCD Model} \label{sec:2_HQCD_Model}

In this section, we introduce the bottom-up holographic QCD model used in this paper to describe massless pions, as well as a tower of massive vector and axial vector mesons. The holographic realisation of these mesons is in terms of components of a five-dimensional gauge field, whose Kaluza--Klein (KK) decomposition includes a massless scalar interpreted as the pion, together with a tower of massive vector and axial vector fields, the lightest of which corresponds to the $\rho$-meson.\,\,For our purposes, the only necessary aspect to compute from the model is the asymptotic UV-behaviour of the pion and $\rho$-meson wavefunctions.

The holographic QCD model we consider is similar to the models in references \cite{PhysRevD.69.065020, Sakai:2004cn}, which aim to describe QCD with $N_f$ massless quarks.\,\,We assume that the low-energy effective theory of our mesons is a U$(N_f)$ gauge theory in five-dimensional AAdS space, described by the following action:
\begin{align} \label{eq:2_HQCD_Action}
    S &\sim -\frac{1}{2} \int d^{4}x dw \sqrt{-g} \, \text{tr} \Big[g^{MN} g^{PQ} F_{MP} F_{NQ}\Big] \nonumber \\
    &\sim -\int d^{4}x dw \sqrt{-g} \, \text{tr} \bigg[\frac{h(w)}{2} \, \eta^{\mu\nu} \eta^{\rho\sigma} F_{\mu\rho} F_{\nu\sigma} + k(w) \, \eta^{\mu\nu} F_{\mu w} F_{\nu w}\bigg],
\end{align}
where $F_{MN}$ is the field strength tensor of the U$(N_f)$ gauge field $A_M$, the coordinates $x^M = (x^{\mu}, w)$ parameterise the bulk space, and recall that $M = 0, \ldots, 4$ and $\mu = 0, \ldots, 3$. We also assume that the metric in our model can be expressed by:
\begin{equation} \label{eq:2_HQCD_Metric}
    ds^2 = g_{MN} dx^M dx^N = a(w) \, \eta_{\mu\nu} dx^{\mu} dx^{\nu} + b(w) \, dw^{2}, \quad -\infty < w < +\infty.
\end{equation}
The functions $h(w)$ and $k(w)$ in the action \eqref{eq:2_HQCD_Action} are defined as follows:
\begin{equation}
    h(w) = \frac{1}{a^2(w)}, \qquad k(w) = \frac{1}{a(w) \, b(w)}.
\end{equation}
In addition, we assume that the metric \eqref{eq:2_HQCD_Metric} approaches the AdS metric as $w \to \pm \infty$, and that the functions $a(w)$ and $b(w)$ are non-singular, positive-valued, and symmetric under the transformation $w \to -w$ for all values of $w$.\,\,This setup is known to be consistent with confinement and chiral symmetry breaking in four-dimensional QCD \cite{PhysRevD.69.065020, Sakai:2004cn}.

More precisely, the action \eqref{eq:2_HQCD_Action} should be interpreted as the kinetic term for massless open string states, while the complete action of our theory is defined to include an infinite tower of excited massive string states, which UV-complete the action \eqref{eq:2_HQCD_Action} (this draws on the understanding found in reference \cite{Imoto:2010ef}).

It will be useful later to know the asymptotic behaviour of the functions $a(w)$~and~$b(w)$, as well as that of the volume element $\sqrt{-g}$, as $w \to \pm \infty$ (i.e.\,\,in the UV-region):
\begin{equation} \label{eq:2_HQCD_UV}
   a(w) \sim \frac{w^2}{R_5^2}, \qquad b(w) \sim \frac{R_5^2}{w^2}, \qquad \sqrt{-g} \sim \frac{|w|^3}{R_5^3},
\end{equation}
where $R_5$ is the radius of curvature of the five-dimensional AAdS space in the UV-region.

To recover four-dimensional physics from the action \eqref{eq:2_HQCD_Action}, we expand $A_M$ in terms of the complete sets of modes $\{\psi_{n}(w)\}_{n \geq 1}$ and $\{\phi_{n}(w)\}_{n \geq 0}$:
\begin{equation} \label{eq:2_Gauge_Exp}
    A_{\mu}(x^{\mu}, w) = \sum_{n=1}^{\infty} B_{\mu}^{(n)}(x^{\mu}) \psi_{n}(w), \quad A_{w}(x^{\mu}, w) = {\varphi}^{(0)}(x^{\mu}) \phi_{0}(w) + \sum_{n=1}^{\infty} \varphi^{(n)}(x^{\mu}) \phi_{n}(w).
\end{equation}
As the lightest vector meson state, $B_{\mu}^{(1)}$ will be identified with the $\rho$-meson field, while $B_{\mu}^{(n)}$ for $n > 1$ with a tower of heavier vector and axial vector meson fields.\,\,The zero mode $\varphi^{(0)}$ will be a massless pion field, identified with the NG boson associated with global chiral symmetry breaking.\footnote{The interpretation of the modes $\varphi^{(0)}$ and $B_{\mu}^{(1)}$ as the pion and $\rho$-meson, respectively, can be shown to be further supported by their parity (see also reference \cite{Sakai:2004cn}).}\,\,The other fields $\varphi^{(n)}$ for $n \geq 1$ will denote fictitious NG bosons that will be absorbed by the fields $B_{\mu}^{(n)}$ via the Stueckelberg mechanism, making the fields $B_{\mu}^{(n)}$ massive in the four-dimensional theory.

Although we could choose to write $A_M$ in a gauge that does not include the fictitious NG bosons, we have decided to retain them, as they will play an essential role later when we use the NG boson equivalence theorem in \hyperref[sec:4_Scaling_Laws]{Section 4}.

To obtain a canonically normalised action for the massless pion $\varphi^{(0)}$ and the massive mesons $B_{\mu}^{(n)}$, we impose the following orthonormality conditions on $\psi_{n}(w)$ and $\phi_{n}(w)$:
\begin{align} \label{eq:2_Orth_Eq}
    \int dw \sqrt{-g} \, h(w) \psi_{n}(w) \psi_{m}(w) = \delta_{nm}, \qquad \int dw \sqrt{-g} \, k(w) \phi_{n}(w) \phi_{m}(w) = \delta_{nm}.
\end{align}
In addition, $\psi_{n}(w)$ and $\phi_{n}(w)$ for $n \geq 1$ are chosen to satisfy the following equations:
\begin{equation} \label{eq:2_Wave_Eq}
    \frac{1}{\sqrt{-g} \, h(w)} \partial_w \Big(\sqrt{-g} \, k(w) \partial_w \psi_{n}(w)\Big) = -m_n^2 \psi_{n}(w), \qquad \partial_w \psi_{n}(w) = m_n \phi_{n}(w),
\end{equation}
where $m_n^2$ are the eigenvalues of the eigenequation written above on the left.\,\,For $n \geq 1$, it can be shown that $m_n^2$ are positive and correspond to the masses of the mesons $B_{\mu}^{(n)}$. Meanwhile, $\phi_{0}(w)$ is chosen to be proportional to $1 / (\sqrt{-g} \, k(w))$, ensuring that it satisfies:
\begin{equation} \label{eq:2_Wave_Eq_Pion}
    \partial_{w} \Big(\sqrt{-g} \, k(w) \phi_{0}(w)\Big) = 0.
\end{equation}

Given equations \eqref{eq:2_Gauge_Exp}--\eqref{eq:2_Wave_Eq_Pion}, the action \eqref{eq:2_HQCD_Action} reads:
\begin{align} \label{eq:2_HQCD_4dAction}
    S \sim -\int d^{4}x \, \text{tr} \bigg[\partial_{\mu} \varphi^{(0)} \partial^{\mu} \varphi^{(0)} + \sum_{n=1}^{\infty} \frac{1}{2} \Big(&\partial_{\mu} B_{\nu}^{(n)} - \partial_{\nu} B_{\mu}^{(n)}\Big)^2 \nonumber \\
    &+ \sum_{n=1}^{\infty} m_n^2 \Big(B_{\mu}^{(n)} - \frac{1}{m_n} \partial_{\mu} \varphi^{(n)}\Big)^2 + \cdots\bigg],
\end{align}
where the dots above represent interaction terms.\,\,In the case where $N_f = 1$, the fictitious NG bosons $\varphi^{(n)}$ for $n \geq 1$ will always appear in the combination $B_{\mu}^{(n)} - \frac{1}{m_n} \partial_{\mu} \varphi^{(n)}$, and thus can be absorbed via the Stueckelberg mechanism.\,\,More generally, however, it can be shown that these fictitious NG bosons can also be gauged away in the non-abelian case. On the other hand, the pion field $\varphi^{(0)}$ cannot be gauged away (e.g.\,\,see reference \cite{Sakai:2004cn}).

The asymptotic behaviour of the pion and $\rho$-meson wavefunctions in the UV-region can be determined by inserting an ansatz of the form $w^{c}$ (with $c$ being an arbitrary constant) into equations \eqref{eq:2_Wave_Eq} and \eqref{eq:2_Wave_Eq_Pion} and requiring that it vanishes in the limit $w \to \pm \infty$, which is necessary to obtain the finite four-dimensional action \eqref{eq:2_HQCD_4dAction}.\,\,The pion wavefunction, due~to the simplicity of equation \eqref{eq:2_Wave_Eq_Pion}, can be expressed as follows for a specific metric:
\begin{equation} \label{eq:wavefunction_pi}
    \psi_{\pi}(w) \coloneqq \phi_{0}(w) = \frac{1}{\sqrt{-g} \, k(w)} \sim \frac{R_5^3}{|w|^3},
\end{equation}
where, in the final expression above, we have presented the leading term in the UV-region. The asymptotic behaviour of the $\rho$-meson wavefunction in the UV-region is:
\begin{equation} \label{eq:wavefunction_rho}
    \psi_{\rho}(w) \coloneqq \psi_{1}(w) \sim \frac{R_5^2}{w^2}.
\end{equation}
In equations \eqref{eq:wavefunction_pi} and \eqref{eq:wavefunction_rho}, we have introduced a new notation for the pion and $\rho$-meson wavefunctions and omitted writing explicit normalisation factors.
\section{Generalising the Polchinski--Strassler Proposal} \label{sec:3_PS_Prop}

In this section, after reviewing the original Polchinski--Strassler proposal \cite{Polchinski:2001tt} in \hyperref[sec:3_1_PS_Original]{Section 3.1}, we generalise it in \hyperref[sec:3_2_PS_General]{Section 3.2}, and conclude with an example of our generalised proposal in~\hyperref[sec:3_3_PS_GenEx]{Section 3.3}.\,\,The main difference from the original proposal is that our proposal partially accounts for the bulk space momenta transverse to the momenta in the dual field theory. However, for pion scattering amplitudes, note that our generalised proposal will reduce to an application of the original proposal in the context of our holographic QCD model.
\subsection{Brief Review of the Original Proposal} \label{sec:3_1_PS_Original}

The original Polchinski--Strassler proposal \cite{Polchinski:2001tt} argues that the high-energy fixed-angle limit of an $n$-glueball scattering amplitude $\mathcal{A}_n$ in a four-dimensional confining gauge theory, holographically dual to a superstring theory in five-dimensional AdS space with a cut-off, can be obtained from a superstring amplitude in five-dimensional Minkowski~space:
\begin{equation} \label{eq:3_PS_S1}
    \mathcal{A}_n \Big(k_{\mu}^{(i)}\Big) = \int dw \sqrt{-g} \times \widetilde{\mathcal S}_n \Big(k_{\mu}^{(i)}\Big) \times \prod_{i=1}^n \psi^{(i)}(w).
\end{equation}
In the previous equation, $k_{\mu}^{(i)}$ for $i = 1, \ldots, n$ are the momenta of the scattered states in four~dimensions, $\psi^{(i)}(w)$ are the wavefunctions of the scattered states, and $\widetilde{\mathcal S}_n$ is obtained from an $n$-superstring scattering amplitude $\mathcal{S}_n$ in five-dimensional Minkowski space inside which the Minkowski metric was replaced with the AdS metric.\,\,In the original proposal~\cite{Polchinski:2001tt}, as~mentioned before, the bulk space was a five-dimensional hard-wall AdS space.\,\,This space can be obtained from equation \eqref{eq:2_HQCD_Metric} by setting $a(w) = 1 / b(w) = w^2 / R_5^2$ and restricting $w$ to $w_{\text IR} < w < \infty$, where $w_{\text IR} > 0$ serves as an IR cut-off.

Next, notice that $\widetilde{\mathcal S}_n$ in the original proposal \eqref{eq:3_PS_S1} does not depend on the momenta of the scattered strings along the $w$-direction, denoted in this context by $k_w^{(i)}$.\,\,This is because reference~\cite{Polchinski:2001tt} argued that $k_w^{(i)}$ can be neglected whenever $\alp/R_5^2 \ll 1$ (this claim will be clarified in the next subsection).\,\,However, as we will show in \hyperref[sec:4_Scaling_Laws]{Section 4}, $k_w^{(i)}$ is essential for obtaining $\rho$-meson scattering amplitudes that are consistent with the NG boson equivalence theorem in our holographic QCD model and thus cannot be completely neglected.

For completeness, we now provide an example of $\widetilde{\mathcal S}_n$ from reference \cite{Polchinski:2001tt}.\,\,In that work, the authors mainly focused on $n$-point scalar glueball scattering.\,\,In holography, glueballs correspond to closed string states, with the scalar glueballs in reference \cite{Polchinski:2001tt} identified with the dilaton.\,\,Hence, for example, in the case of 2-to-2 scalar glueball scattering, the dual superstring scattering amplitude at tree level in the original proposal \eqref{eq:3_PS_S1} is given by:
\begin{equation}
    \widetilde{\mathcal S}_4 \Big(k_{\mu}^{(i)}\Big) \propto g_s^4 \, \frac{\Gamma \big(-\frac{\alp \Tilde{s}}{4}\big) \Gamma \big(-\frac{\alp \Tilde{t}}{4}\big) \Gamma \big(-\frac{\alp \Tilde{u}}{4}\big)}{\Gamma \big(1+\frac{\alp \Tilde{s}}{4}\big) \Gamma \big(1+\frac{\alp \Tilde{t}}{4}\big) \Gamma \big(1+\frac{\alp \Tilde{u}}{4}\big)} \widetilde{\Kin}_4 \Big(k_{\mu}^{(i)}\Big),
\end{equation}
where $g_s$ is the string coupling constant, $\alp$ is the Regge slope, $\widetilde{\Kin}_4$ is a kinematical factor that depends on the four-dimensional momenta $k_{\mu}^{(i)}$, and $\tilde{s}$ is defined as follows:
\begin{equation} \label{eq:3_Mandelstam_Warped}
    \tilde{s} = - g^{\mu\nu} \Big(k_{\mu}^{(1)}+k_{\mu}^{(2)}\Big) \Big(k_{\nu}^{(1)}+k_{\nu}^{(2)}\Big) = -\frac{\eta^{\mu\nu}}{a(w)} \Big(k_{\mu}^{(1)}+k_{\mu}^{(2)}\Big) \Big(k_{\nu}^{(1)}+k_{\nu}^{(2)}\Big) = \frac{s}{a(w)},
\end{equation}
with similar definitions for the variables $\tilde{t}$ and $\tilde{u}$.\,\,Here, note that a tilde above a quantity (e.g.\,\,$\widetilde{\Kin}_4$ or $\tilde{s}$) indicates that contractions of indices of momenta and polarisations inside it are performed using the AdS metric instead of the Minkowski metric.

Moreover, reference \cite{Polchinski:2001tt} demonstrated that, starting from $n$-point dilaton scattering in five-dimensional hard-wall AdS space, the original proposal \eqref{eq:3_PS_S1} predicts the following behaviour for scalar glueball scattering amplitudes in the high-energy fixed-angle limit:
\begin{equation} \label{eq:3_PS_Result}
    \mathcal{A}_n \sim s^{2-\frac{\Delta}{2}},
\end{equation}
where $\Delta = \sum_{i=1}^{n} \Delta_i$ with $\Delta_i$ denoting the conformal dimension of the lowest-dimension operator that can create the $i^{\text{th}}$ scattered state.\,\,The authors also argued that $\Delta$ should be interpreted as the minimum number of hard constituents involved in the scattering process. Given this interpretation, the result \eqref{eq:3_PS_Result} agrees with the constituent counting rule \eqref{eq:1_Scaling_Rule}.

The authors in reference \cite{Polchinski:2001tt} further asserted that their result \eqref{eq:3_PS_Result} can be generalised to the scattering of states with four-dimensional spin, in agreement with the constituent counting rule.\,\,However, as we will show, the original proposal cannot be naively extended to the latter problem and requires modification before it can be applied more generally.

In the next subsection, we will present our generalisation of the original proposal \eqref{eq:3_PS_S1}. We will explain exactly why the original proposal requires modification later in \hyperref[sec:4_2_PS_Vectors]{Section 4.2}.
\subsection{Statement of the Generalised Proposal} \label{sec:3_2_PS_General}

In the context of our holographic QCD model from \hyperref[sec:2_HQCD_Model]{Section 2}, we propose that an $n$-point scattering amplitude $\mathcal{A}_n$ of four-dimensional mesons, arising from a U$(N_f)$ gauge field in five-dimensional AAdS space, in the high-energy fixed-angle limit is given by:
\begin{equation} \label{eq:3_PS_Gen}
    \mathcal{A}_n \Big(k_{\mu}^{(i)}, \zeta_{\mu}^{(i)}\Big) = \int dw \sqrt{-g} \times \widetilde{\mathcal S}_n \Big(p_M^{(i)}, \zeta_M^{(i)}\Big) \times \prod_{i=1}^n \psi^{(i)}(w),
\end{equation}
where $k_{\mu}^{(i)}$ for $i = 1, \ldots, n$ are the momenta of the scattered mesons, which include pions, as well as vector and axial vector mesons.\footnote{From now on, we will primarily focus on the pion and $\rho$-meson.\,\,However, our findings will also directly apply to the other vector and axial vector mesons captured by the holographic QCD model from \hyperref[sec:2_HQCD_Model]{Section 2}.}\,\,The polarisations $\zeta_{\mu}^{(i)}$ are reserved for the vector and axial vector mesons.\,\,Here, a key difference from the original proposal \eqref{eq:3_PS_S1} is that $\widetilde{\mathcal S}_n$ is now a differential operator acting on the wavefunctions $\psi^{(i)}(w)$ of the scattered mesons.

More precisely, $\widetilde{\mathcal S}_n$ is constructed from an open superstring scattering amplitude~$\mathcal{S}_n$ for U$(N_f)$ gauge bosons, with momenta $p_M^{(i)}$ and polarisations $\zeta_M^{(i)}$, in five-dimensional Minkowski space.\,\,For 2-to-2 scattering at tree level, with the Chan--Paton factors omitted (a practice we adopt henceforth, unless specified otherwise), $\mathcal{S}_4$ is given by \cite{Green:1987sp, Polchinski:1998rr}:
\begin{equation} \label{eq:3_S4}
    \mathcal{S}_4 \Big(p_M^{(i)}, \zeta_M^{(i)}\Big) = F_4^{M_1 N_1 \cdots M_4 N_4} \Big(p_M^{(i)} \sqrt{\alp}\Big) \, M_{M_1 N_1}^{(1)} \cdots M_{M_4 N_4}^{(4)},
\end{equation}
where:
\begin{equation} \label{eq:3_F4}
    F_4^{M_1 N_1 \cdots M_4 N_4} \Big(p_M^{(i)} \sqrt{\alp}\Big) = -\frac{g_s^2}{2} \, \frac{\Gamma \big(-\alp s\big) \, \Gamma \big(-\alp t\big)}{\Gamma \big( 1-\alp s-\alp t\big)} \, t^{M_1 N_1 \cdots M_4 N_4}.
\end{equation}
Here, $s$ and $t$ are the Mandelstam variables in five-dimensional Minkowski space, and $t^{\cdots}$ is a tensor constructed solely from combinations of the five-dimensional Minkowski metric (for an explicit definition, see references \cite{Green:1987sp, Polchinski:1998rr}).\,\,In addition, the tensor $t^{\cdots}$ is antisymmetric under the exchange of its indices $M_i \leftrightarrow N_i$ and symmetric under the exchange of any pair of particles $(M_i N_i) \leftrightarrow (M_j N_j)$ for $i, \, j = 1, \ldots, 4$.\,\,Moreover, $M_{MN}^{(i)}$ is expressed as:
\begin{equation}
    M_{MN}^{(i)} = p_M^{(i)} \zeta_N^{(i)} - p_N^{(i)} \zeta_M^{(i)}.
\end{equation}

Returning to our proposal \eqref{eq:3_PS_Gen} for any $n$, we now provide a precise prescription for the case where $N_f = 1$.\footnote{Note that $n$-point scattering amplitudes of U$(1)$ gauge bosons vanish when $n$ is odd, as a consequence of charge conjugation symmetry (for additional details, see Furry's theorem on page 318 of reference \cite{Peskin:1995ev}).}\,\,We focus on $N_f = 1$ because the $n$-photon superstring scattering amplitude can be expressed in terms of $M_{MN}^{(i)}$ \cite{Andreev:1988bz}.\,\,For 4-meson scattering, our proposal also holds for any $N_f$, as the 4-point open superstring scattering amplitude can likewise be written using $M_{MN}^{(i)}$, as shown in equation \eqref{eq:3_S4}.\,\,The reason that prevents us from giving a precise prescription for the case with $N_f > 1$ and any $n$ will be discussed in \hyperref[sec:5_Conclusions]{Section 5}.

In this context, $\mathcal{S}_n$ is given by:
\begin{equation} \label{eq:S_n_generic}
    \mathcal{S}_n \Big(p_M^{(i)}, \zeta_M^{(i)}\Big) = F_n^{M_1 N_1 \cdots M_n N_n} \Big(p_M^{(i)} \sqrt{\alp}\Big) \, M_{M_1 N_1}^{(1)} \cdots M_{M_n N_n}^{(n)}.
\end{equation}
In the previous equation, $F_n^{\cdots}$ is a tensor formed from combinations of the five-dimensional Minkowski metric and momenta $p_M^{(i)}$ (see equation \eqref{eq:3_F4} for an example).

Following the preceding discussion, in our proposal \eqref{eq:3_PS_Gen}, $\widetilde{\mathcal S}_n$ for $n$-point scattering of U$(1)$ gauge bosons and 4-point scattering of U$(N_f)$ gauge bosons is given by:
\begin{equation} \label{eq:3_GenPS_FK}
    \widetilde{\mathcal S}_n \Big(p_M^{(i)}, \zeta_M^{(i)}\Big) = \widetilde{F}_n^{M_1 N_1 \cdots M_n N_n} \Big(p_{\mu}^{(i)} \sqrt{\alp}\Big) \, \widetilde{M}_{M_1 N_1}^{(1)} \cdots \widetilde{M}_{M_n N_n}^{(n)},
\end{equation}
where $\widetilde{F}_n^{\cdots}$ and $\widetilde{M}_{MN}^{(i)}$ are to be interpreted as $F_n^{\cdots}$ and $M_{MN}^{(i)}$, subject to the following replacements (\textbf{i}), (\textbf{ii}), and (\textbf{iii}).\,\,In analogy to this, in our generalised proposal, any other quantity with a tilde above it (e.g.\,\,$\widetilde{s}$) is also subject to the replacements (\textbf{i}), (\textbf{ii}), and (\textbf{iii}). Before we proceed, notice that the components $p_w^{(i)}$ are omitted in $\widetilde{F}_n^{\cdots}$ in equation \eqref{eq:3_GenPS_FK} (a justification for this will be provided by the end of this subsection).
\begin{enumerate} [label=(\textbf{\roman*})]
    \item The five-dimensional Minkowski metric $\eta_{MN}$ is replaced with the AAdS metric $g_{MN}$ from equation \eqref{eq:2_HQCD_Metric}:
    \begin{equation} \label{eq:3_Replace_Metric}
    \eta_{MN} \to g_{MN}.
    \end{equation}
    \item The four-dimensional momenta $p_{\mu}^{(i)}$ are replaced with the momenta $k_{\mu}^{(i)}$, while $p_w^{(i)}$ are replaced with covariant derivatives $-i \nabla_w^{(i)}$ that act only on the wavefunctions $\psi^{(i)}(w)$ with the same index $(i)$:
    \begin{equation} \label{eq:3_Replace_Mom}
    p_{\mu}^{(i)} \to k_{\mu}^{(i)}, \qquad p_w^{(i)} \to -i \nabla_w^{(i)},
    \end{equation}
    with $\nabla_w^{(i)}$ and $\nabla_w^{(j)}$ commuting for $i \neq j$.\,\,Moreover, note that in equation \eqref{eq:3_GenPS_FK}, only the factors $\widetilde{M}_{MN}^{(i)}$ act as differential operators, and that in these factors, $p_w^{(i)}$ can be replaced with $-i \partial_w^{(i)}$, as the contributions from the Christoffel symbols cancel.
    \item The polarisations of the vector and axial vector mesons $B_{\mu}^{(n)}$ are set to:
    \begin{equation} \label{eq:3_Replace_Vec}
    \zeta_M^{(i)} \to \big(\zeta_{\mu}^{(i)}, 0\big),
    \end{equation}
    while the ``polarisations'' for the pion $\varphi^{(0)}$ and fictitious NG bosons $\varphi^{(n)}$ are set to:
    \begin{equation} \label{eq:3_Replace_Scalar}
    \zeta_M^{(i)} \to \big(0, 0, 0, 0, \zeta_w^{(i)}\big).
    \end{equation}
    In addition, the wavefunctions $\psi^{(i)}(w)$ are replaced according to:
    \begin{align} \label{eq:3_Replace}
    \zeta_{\mu}^{(i)} \psi^{(i)}(w) \rightarrow \zeta_{\mu}^{(i)} \psi_{n_i}(w), \qquad
    \zeta_w^{(i)} \psi^{(i)}(w) \rightarrow \phi_{n_i}(w).
\end{align}
\end{enumerate}

Now, let us explain why the components $p_w^{(i)}$ can be neglected in $\widetilde{F}_n^{\cdots}$.\,\,First, note that the high-energy fixed-angle limit of an $n$-point scattering amplitude refers to the limit in which the absolute values of all generalised Mandelstam variables $s_{ij} = -\big( k^{(i)} + k^{(j)} \big)^2$ for $i, \, j = 1, \ldots, n$ and $i \neq j$ simultaneously approach infinity, while their ratios, corresponding to the scattering angles $\theta_i$, remain fixed.\,\,In this limit, the scattering amplitude depends only on $s$ (i.e.\,\,$s_{12}$) and several independent scattering angles $\theta_i$.

In the high-energy fixed-angle limit, the leading-order contribution in $s$ to $\mathcal{A}_n$ comes from the UV-region where $w \to \pm \infty$, since outside this region, $\mathcal{A}_n$ is strongly suppressed by the exponential softness of the superstring scattering amplitude \cite{Polchinski:2001tt}.\,\,Then, focusing on the high-energy fixed-angle limit and the UV-region, we introduce the variable $\tilde{s}$:
\begin{equation} \label{eq:3_Comparison}
    \tilde{s} = -g^{MN} \Big(p_M^{(1)}+p_M^{(2)}\Big) \Big(p_N^{(1)}+p_N^{(2)}\Big) \sim \frac{R_5^2}{w^2} \, s+\frac{w^2}{R_5^2} \, \nabla_w^{(1)} \nabla_w^{(2)},
\end{equation}
where $s$ is the Mandelstam variable in four-dimensional Minkowski space, and in the final expression, we used the identities $p_{\mu}^{(i)} \to k_{\mu}^{(i)}$ and $p_w^{(i)} \to -i \nabla_w^{(i)}$ and assumed that $w \to \pm \infty$.

Given the above discussion, the dominant contribution to the integral in equation~\eqref{eq:3_PS_Gen} comes from the region where $\alp \tilde{s} \sim \mathcal{O}(1)$.\,\,Now, recall from equations \eqref{eq:wavefunction_pi} and \eqref{eq:wavefunction_rho} that in the UV-region, the wavefunctions in our holographic model have power-law behaviour, implying that in this region $\nabla_w^{(i)} \sim 1/w$.\,\,Thus, $\alp \tilde{s}$ simplifies in the UV-region as follows:
\begin{equation} \label{eq:3_SecondT}
    \alp \tilde{s} \sim \frac{1}{w_{*}^2} + \frac{\alp}{R_5^2},
\end{equation}
where $w_{*} \coloneqq w/(R_5 \sqrt{\alp s})$.\,\,Then, requiring that $\alp \tilde{s} \sim \mathcal{O}(1)$, we find that the leading-order contribution comes from $1 / w_{*}^2 \sim \mathcal{O}(1) \gg \alp/R_5^2$.\,\,Hence, the contribution from the second term in equation \eqref{eq:3_SecondT} to $\alp \tilde{s}$ can be neglected, assuming that $\alp/R_5^2 \ll 1$.

The discussion above should have clarified why reference \cite{Polchinski:2001tt} argued that $p_w^{(i)}$ can be neglected in $\widetilde{\mathcal S}_n$.\,\,However, we emphasise that this is not always true.\,\,The components $p_w^{(i)}$ can only be neglected in $\widetilde{F}_n^{\cdots}$, while they must be kept in $\widetilde{M}_{MN}^{(i)}$.\,\,This is because, as we will see in \hyperref[sec:4_Scaling_Laws]{Section 4}, for processes with $\rho$-mesons, the leading term comes from $p_w^{(i)}$ in $\widetilde{M}_{\mu w}^{(i)}$. On the other hand, notice that for scattering amplitudes involving only pions or fictitious NG bosons, the form of $\widetilde{M}_{MN}^{(i)}$ does not allow for contributions from the components $p_w^{(i)}$ in these factors, given the ``polarisations'' \eqref{eq:3_Replace_Scalar}.\,\,Thus, for scattering amplitudes involving pions or fictitious NG bosons, the contributions from the components $p_w^{(i)}$ can be completely neglected in the entire scattering amplitude $\widetilde{\mathcal{S}}_n$.

Even though the dependence of $\widetilde{F}_n^{\cdots}$ on $p_M^{(i)}$ extends beyond the Mandelstam variables (e.g.\,\,momenta $p_M^{(i)}$ with uncontracted indices may appear in $\widetilde{F}_n^{\cdots}$), a similar argument to the~one above shows that $p_w^{(i)}$ can be neglected in $\widetilde{F}_n^{\cdots}$, provided that the leading-order term is~non-zero.\,\,In \hyperref[app:C_pw_zero]{Appendix C}, we show that the components $F_n^{\mu_1 w \cdots \mu_n w}$ are non-vanishing when $p_w^{(i)}=0$ for all $i$, thereby completing our argument for neglecting $p_w^{(i)}$ in $\widetilde{F}_n^{\cdots}$.
\subsection{A Simple Example of the Generalised Proposal} \label{sec:3_3_PS_GenEx}

\begin{figure}[ht] \label{fig:Toy_Diagram}
    \centering
    \includegraphics[scale = 1]{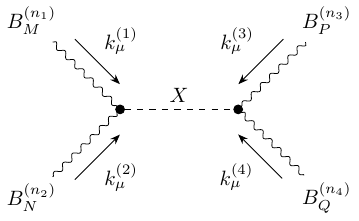}
    \caption{An $s$-channel Feynman diagram contributing to the $2$-to-$2$ scattering of modes $B_{\mu}^{(n_{i})}$ and $B_w^{(n_{i})} \coloneqq \varphi^{(n_i)}$ with momenta $k_{\mu}^{(i)}$ for $i = 1, \ldots, 4$.}
\end{figure}

To motivate our proposal \eqref{eq:3_PS_Gen}, we examine 2-to-2 scattering amplitudes in a toy model described by a five-dimensional action in AAdS space that, in addition to the kinetic term of the gauge field $A_M$ from equation \eqref{eq:2_HQCD_Action}, includes the following terms:
\begin{equation} \label{eq:3_Action_Toy}
    S = \int d^{4}x dw \sqrt{-g} \Big[g^{MN} \partial_M X \partial_N X - M^2 X^2 + X g^{MN} g^{PQ} F_{MP} F_{NQ}\Big],
\end{equation}
where $X$ is a massive real scalar field serving as the exchanged mode in the scattering amplitudes that we will study (see \hyperref[fig:Toy_Diagram]{Figure 1} on the next page), and the metric is given by equation \eqref{eq:2_HQCD_Metric}.\,\,For simplicity, we consider the case with $N_f = 1$.

To compute the scattering amplitude presented in \hyperref[fig:Toy_Diagram]{Figure 1}, we assume that $A_{M}$ takes the following form:
\begin{align}
    A_{\mu}(x^{\mu}, w) &= B_{\mu}^{(n)}(x^{\mu}) \, \psi_{n}(w) = \zeta_{\mu} \psi_{n}(w) \, e^{i k_{\mu} x^{\mu}}, \label{eq:3_Given1}\\
    A_{w}(x^{\mu}, w) &= \varphi^{(n)}(x^{\mu}) \, \phi_{n}(w) = \phi_{n}(w) \, e^{i k_{\mu} x^{\mu}}, \label{eq:3_Given2}
\end{align}
where $k_{\mu}$ and $\zeta_{\mu}$ are the on-shell four-dimensional momenta and polarisations, respectively, while the wavefunctions $\psi_{n}(w)$ and $\phi_{n}(w)$ were introduced earlier in \hyperref[sec:2_HQCD_Model]{Section 2}.

Now, given equations \eqref{eq:3_Given1} and \eqref{eq:3_Given2}, the field strength tensor $F_{MN}$ can be expressed as $i \widetilde{M}_{MN} \, \psi(w) \exp{(i k_{\mu} x^{\mu})}$, where $\widetilde{M}_{MN}$ is a differential operator defined as:
\begin{equation} \label{MMN}
    \widetilde{M}_{MN} = -i \big(\zeta_N \partial_M - \zeta_M \partial_N \big),
\end{equation}
and the wavefunction $\psi(w)$ must be replaced as in equation \eqref{eq:3_Replace}.\,\,The interaction term in the action \eqref{eq:3_Action_Toy} corresponding to the three-point vertex on the left side of the diagram in \hyperref[fig:Toy_Diagram]{Figure 1} can then be written as follows:
\begin{align}
    S &\sim \int d^{4}x dw \sqrt{-g} \, X g^{MN} g^{PQ} \widetilde{M}_{MP}^{(1)} \widetilde{M}_{NQ}^{(2)} \, \prod_{i=1}^{2} \psi^{(i)}(w) \, e^{i k_{\mu}^{(i)} x^{\mu}} \nonumber \\
    &\sim \int d^{4}x dw \sqrt{-g} \, X J^{(1, 2)}(x^{\mu}, w).
\end{align}
In the equation above, we have implicitly defined the source term $J^{(1, 2)}(x^{\mu}, w)$:
\begin{equation}
    J^{(1, 2)}(x^{\mu}, w) = g^{MN} g^{PQ} \widetilde{M}_{MP}^{(1)} \widetilde{M}_{NQ}^{(2)} \, \prod_{i=1}^{2} \psi^{(i)}(w) \, e^{i k_{\mu}^{(i)} x^{\mu}}.
\end{equation}
Note that the index $(i)$ on $\widetilde{M}_{MN}^{(i)}$ and $\psi^{(i)}(w) \exp(i k_{\mu}^{(i)} x^{\mu})$ indicates that the combination $\psi^{(i)}(w) \exp(i k_{\mu}^{(i)} x^{\mu})$ represents the wavefunction of the $i^{\text{th}}$ scattered meson, and that $\widetilde{M}_{MN}^{(i)}$ acts exclusively on it.

The propagator for the exchanged five-dimensional massive real scalar field $X$ can be formally derived from its Green's function $G\big(x^M; y^M\big)$:
\begin{align} \label{eq:3_RHS}
    G \big(x^M; y^M\big) = \frac{1}{\sqrt{-g}} \, \frac{1}{\nabla^2 + M^2} \, \delta^{(5)} \big(x^M - y^M\big).
\end{align}
Furthermore, when the propagator $G(x^{M}; y^{M})$ is used to compute the diagram in \hyperref[fig:Toy_Diagram]{Figure 1}, we can apply the same reasoning as in the previous subsection to neglect the $\nabla_w^2$-component on the right-hand side of equation \eqref{eq:3_RHS}, assuming that $M^2 R_5^2 \gg 1$:
\begin{align}
    G \big(x^M; y^M\big) \sim \frac{1}{\sqrt{-g}} \, \frac{1}{-\tilde{s} + M^2} \, \delta^{(5)} \big(x^M - y^M\big),
\end{align}
where we introduced the rescaled Mandelstam variable $\tilde{s}$:
\begin{equation}
    \tilde{s} = -g^{\mu \nu} \Big(k_{\mu}^{(1)}+k_{\mu}^{(2)}\Big)\Big(k_{\nu}^{(1)}+k_{\nu}^{(2)}\Big).
\end{equation}
Here, note that $p_w^{(i)}$ was omitted from the definition of $\tilde{s}$ compared to equation \eqref{eq:3_Comparison}.

Thus, an $s$-channel contribution to the $2$-to-$2$ tree level scattering amplitude presented in \hyperref[fig:Toy_Diagram]{Figure 1} can be expressed formally as follows:
\begin{align} \label{eq:3_manifest}
    \mathcal{A} &= \int d^{5}x d^{5}y \sqrt{-g \big(x^M\big)} \sqrt{-g \big(y^M\big)} \, J^{(1, 2)} \big(x^M\big) G \big(x^M; y^M\big) J^{(3, 4)} \big(y^M\big) \nonumber \\
    &\sim \int d^{4}x dw \sqrt{-g} \, J^{(1, 2)}(x^{\mu}, w) \, \frac{1}{-\tilde{s} + M^{2}} \, J^{(3, 4)}(x^{\mu}, w) \nonumber \\
    &\sim \delta^{(4)} \Big( {\textstyle \sum_{i=1}^4 k_{\mu}^{(i)}} \Big) \int dw \sqrt{-g} \times \widetilde{\mathcal S} \Big(p_M^{(i)}, \zeta_M^{(i)}\Big) \times \prod_{i=1}^{4} \psi^{(i)}(w),
\end{align}
where:
\begin{equation}
    \widetilde{\mathcal S} \Big(p_M^{(i)}, \zeta_M^{(i)}\Big) = \frac{1}{-\tilde{s} + M^2} \, g^{M_1 M_2} g^{N_1 N_2} g^{M_3 M_4} g^{N_3 N_4} \widetilde{M}_{M_1 N_1}^{(1)} \cdots \widetilde{M}_{M_4 N_4}^{(4)}.
\end{equation}
In summary, the result \eqref{eq:3_manifest} offers a concrete example of our generalised proposal \eqref{eq:3_PS_Gen}.
\section{The Constituent Counting Rule for Meson Scattering} \label{sec:4_Scaling_Laws}

In asymptotically free confining theories in four dimensions (e.g.\,\,in four-dimensional QCD), scattering amplitudes of low-lying hadrons exhibit characteristic scaling behaviour with $s$ in the high-energy fixed-angle limit, as described by the constituent counting rule \cite{Matveev:1973ra, Brodsky:1973kr, Brodsky:1974vy}:
\begin{equation} \label{eq:4_CC_Rule}
    \mathcal{A}(s, \theta_i) \sim s^{2-\frac{m}{2}} f(\theta_i),
\end{equation}
where $m$ is the minimum number of hard constituents involved in the scattering process, and $f(\theta_i)$ captures the dependence of the scattering amplitude on the scattering angles $\theta_i$ (see \hyperref[app:A_Scaling_Rules]{Appendix A} for a brief review of a derivation of the constituent counting rule).

Now, let $m_j$ denote the minimum number of hard constituents in the $j^{\text{th}}$ hadron, then:
\begin{equation}
    m = \sum_{j=1}^n m_j,
\end{equation}
for a scattering process with $n$ hadrons.\,\,In the context of the quark model, $m_j$ represents the number of quarks that make up the $j^{\text{th}}$ hadron.\,\,Thus, the prediction for the scaling of 2-to-2 meson scattering amplitudes with $s$ in the high-energy fixed-angle limit is:
\begin{equation} \label{eq:4_Pred1} 
    \mathcal{A}_{4}(s, \theta) \sim s^{-2} f(\theta),
\end{equation}
while for $n$-meson scattering, we find:
\begin{equation} \label{eq:4_Pred2}
    \mathcal{A}_{n}(s, \theta_i) \sim s^{2-n} f(\theta_i).
\end{equation}
\clearpage

In this section, we study the scaling with $s$ of pion and $\rho$-meson scattering amplitudes in the high-energy fixed-angle limit, in the context of our holographic model from \hyperref[sec:2_HQCD_Model]{Section 2}. Our objective is to reproduce equations \eqref{eq:4_Pred1} and \eqref{eq:4_Pred2}.\,\,Before we proceed, recall that our holographic model contains an infinite tower of excited massive string states, which must~be exchanged during meson scattering at high energies.\,\,To construct scattering amplitudes for pions and $\rho$-mesons that include these exchanged modes, we naively apply the original proposal~\cite{Polchinski:2001tt} in \hyperref[sec:4_1_Naive_Counting]{Section 4.1} and \hyperref[sec:4_2_PS_Vectors]{Section 4.2}, and use our generalised proposal in \hyperref[sec:4_3_Scaling_Corrected]{Section 4.3}.
\subsection{Naive Power Counting} \label{sec:4_1_Naive_Counting}

Naively applying the original proposal \cite{Polchinski:2001tt}, reviewed earlier in \hyperref[sec:3_1_PS_Original]{Section 3.1}, to an $n$-meson scattering amplitude $\mathcal{A}_n$ in our holographic QCD model, we obtain the following expression in the high-energy fixed-angle limit:
\begin{equation} \label{eq:4_A_n_meson}
    \mathcal{A}_n \Big(k_{\mu}^{(i)}, \zeta_{\mu}^{(i)}\Big) = \int dw \sqrt{-g} \times \widetilde{\mathcal S}_n \Big(k_{\mu}^{(i)}, \zeta_M^{(i)}\Big) \times \prod_{i=1}^n \psi^{(i)}(w),
\end{equation}
where $k_w^{(i)}$ are set to zero for all $i$, as instructed by the original proposal.\,\,In the above, the polarisations $\zeta_M^{(i)}$ should be interpreted according to the equations \eqref{eq:3_Replace_Vec} and \eqref{eq:3_Replace_Scalar}. Moreover, recall that the wavefunctions $\psi^{(i)}(w)$ must be replaced as in equation \eqref{eq:3_Replace}.

The 4-point superstring scattering amplitude was explicitly written in equation \eqref{eq:3_GenPS_FK}. However, we will not need to make detailed assumptions about the superstring scattering amplitude to obtain the constituent counting rule through simple power counting analysis.

It suffices to observe that the integral in equation \eqref{eq:4_A_n_meson} is dominated by the region where $w/(R_5 \sqrt{\alp s}) \sim \mathcal{O}(1)$ in the high-energy fixed-angle limit, as discussed in \hyperref[sec:3_2_PS_General]{Section 3.2}. Then, changing the integration variable to the dimensionless variable:
\begin{equation} \label{eq:4_rescaling}
    w_{*} = \frac{w}{R_5} \, \frac{1}{\sqrt{\alp s}},
\end{equation}
should allow us to extract the $s$-dependence by making the following observations.

The superstring amplitude $\widetilde{\mathcal S}_n$ can depend on the dot products of momenta and polarisations: $k^{(i)} \cdot k^{(j)}$, $k^{(i)} \cdot \zeta^{(j)}$, and $\zeta^{(i)} \cdot \zeta^{(j)}$ for $i \neq j$, defined by contraction with the~AAdS metric \eqref{eq:2_HQCD_Metric}.\,\,We will show below that, after rescaling \eqref{eq:4_rescaling}, these are all \emph{effectively} of $\mathcal{O}(1)$.

The simplest are the products $k^{(i)} \cdot k^{(j)}$ directly related to the Mandelstam variables. In the region where $w_{*} \sim \mathcal{O}(1)$, these products behave as follows:
\begin{equation} \label{eq:4_arg}
    -k^{(i)} \cdot k^{(j)} \to -g^{\mu\nu} k_{\mu}^{(i)} k_{\nu}^{(j)} =  -\frac{\eta^{\mu\nu}}{a(w)} \, k_{\mu}^{(i)} k_{\nu}^{(j)} \sim \frac{1}{\alp w_{*}^2} \sim \mathcal{O}(1).
\end{equation}

As for the polarisations, we consider pions and $\rho$-mesons separately.\,\,For the $\rho$-mesons, transverse polarisations lead to scattering amplitudes $\mathcal{A}_n$ that are suppressed in $s$ compared to the constituent counting rule, as the reader can confirm from the following discussion. To proceed, let us focus on longitudinally polarised $\rho$-mesons, where $\zeta_{\mu} \approx k_{\mu}/m_{\rho} \sim \mathcal{O}(\sqrt{s})$. For such longitudinally polarised $\rho$-mesons, the products $k^{(i)} \cdot \zeta^{(j)}$ and $\zeta^{(i)} \cdot \zeta^{(j)}$ are found to be essentially equivalent to the Mandelstam variables, and the previous argument \eqref{eq:4_arg} applies to them as well.

To understand how the polarisation of a $\rho$-meson behaves at high energies, consider a~generic polarisation $\zeta_{\mu}^{\text{r.f.}}$ in the rest frame and boost it along $\vec{k}$ to obtain $\zeta_{\mu}^{\text{l.f.}} \approx \lambda k_{\mu}/m_{\rho}$ in~the lab frame as $|\vec{k}| \to \infty$, where $\lambda \sim \mathcal{O}(1)$ (for additional details, see \hyperref[app:D_zeta_choice]{Appendix D}). Thus,~if we aim to describe a generic polarisation vector that has a non-zero component along the longitudinal direction in the rest frame, we necessarily obtain a ${\cal O}(\sqrt{s})$ contribution in the lab frame.

For the pions and fictitious NG bosons, the polarisation $\zeta_w$ is naively of ${\cal O}(1)$ since it is independent of momentum.\,\,However, when it appears next to the wavefunction $\psi^{(i)}(w)$, as~in equation \eqref{eq:3_Replace}, we can use equations \eqref{eq:2_Wave_Eq}, \eqref{eq:wavefunction_pi}, and the fact that in the UV-region:
\begin{equation} \label{eq:4_SeeAfter}
    \phi_{n_i}(w) \sim \frac{R_5^3}{w^3},
\end{equation}
to conclude that \emph{effectively} $\zeta_w \sim R_5/w$ if we take the wavefunctions entering equation \eqref{eq:4_A_n_meson} to be $\psi^{(i)}(w) \sim R_5^2/w^2$.\,\,Thus, effectively $g^{ww} \zeta_w \zeta_w \sim \mathcal{O}(1)$.

Then, we can extract the scaling with $s$ for all pion and $\rho$-meson scattering amplitudes from the following contributions (\textbf{a}) and (\textbf{b}).
\begin{enumerate} [label=(\textbf{\alph*})]
    \item The integration measure, which in the UV-region becomes:
    \begin{equation}
        dw \sqrt{-g} \sim dw \, \frac{|w|^3}{R_5^3} \sim d{w_{*}} |w_{*}|^3 R_5 \, \big(\alp s\big)^2.
    \end{equation}
    \item The product of wavefunctions, where each wavefunction in the UV-region reads:
    \begin{equation}
        \psi^{(i)}(w) \sim \frac{R_5^2}{w^2} \sim \frac{1}{w_{*}^2} \, \frac{1}{\alp s}.
    \end{equation}
    In total the product of $n$-wavefunctions $\psi^{(i)}(w)$ gives a factor of $s^{-n}$.
\end{enumerate} 

This completes the power counting argument, as we can write:
\begin{equation} \label{eq:4_A_n_pion_rho}
    \mathcal{A}_n \Big(k_{\mu}^{(i)}, \zeta_{\mu}^{(i)}\Big) \sim s^{2-n} \int d{w_{*}} \, \mathcal{F}_{n}\big(w_{*}; \theta_i\big).
\end{equation}
In the above, the $s$-dependence in the high-energy fixed-angle limit is entirely contained in the prefactor $s^{2-n}$.\,\,The function $\mathcal{F}_n$ in the integral above depends on the precise form of the superstring scattering amplitude, wavefunctions, background, and the specific choice of polarisations, which together determine the scattering angle dependence of $\mathcal{A}_n$.

In conclusion, the result \eqref{eq:4_A_n_pion_rho} seems to agree with the constituent counting rule~\eqref{eq:4_Pred2}. However, as we will soon demonstrate, our discussion in this subsection was incomplete, and the original proposal, when applied naively in the context of our holographic QCD model, fails to capture the expected scaling \eqref{eq:4_Pred2} for scattering processes involving $\rho$-mesons.
\subsection{Gauge Invariance in Vector Meson Scattering} \label{sec:4_2_PS_Vectors}

The power counting analysis from the previous subsection does not provide the full picture, even though it appears to produce results in agreement with the constituent counting rule. The crucial point missing from the previous power counting analysis is that the flat-space superstring amplitude $\mathcal{S}_n$ (of massless gauge bosons) is gauge invariant and must satisfy:
\begin{equation} \label{eq:4_Massless_Gauge}
    \mathcal{S}_n \Big(k_M^{(i)}, \zeta_M^{(i)}; \zeta_M^{(1)} \to \lambda_1 k_M^{(1)}\Big) = 0,
\end{equation}
where without loss of generality, we replaced $\zeta_M^{(1)} \to \lambda_1 k_M^{(1)}$.\,\,However, the same applies for $\zeta_M^{(i)} \to \lambda_i k_M^{(i)}$ for any $i = 1, \dots, n$, with each $\lambda_i$ being an arbitrary constant, as the gauge transformation can be performed locally on any asymptotic state.

The simple power counting argument from the previous subsection for the scaling of the scattering amplitudes with $\rho$-mesons assumed that $\rho$-meson polarisations scale as $\sqrt{s}$. However, this applies only for the longitudinal component of the $\rho$-meson polarisations, which is exactly the component for which $\mathcal{S}_n$ vanishes.\,\,Thus, a closer examination reveals that naively applying the original proposal \cite{Polchinski:2001tt} predicts that scattering amplitudes involving $\rho$-mesons are suppressed by $1/s$ relative to those involving only pions, or, in other words, it predicts that the leading term of order $s^{2-n}$ vanishes.

This prediction is problematic, as there is no a priori reason in field theory to expect that the $\rho$-meson scattering amplitudes should be suppressed compared to the prediction of the constituent counting rule.\,\,More importantly, we find that this result is inconsistent with our holographic QCD model, where gauge invariance should manifest itself through the NG boson equivalence theorem.

Recall from \hyperref[sec:2_HQCD_Model]{Section 2} that in our holographic model, the massive four-dimensional $\rho$-meson and its tower of excited KK modes $B_{\mu}^{(n)}$ arise from a massless five-dimensional gauge field $A_M$.\,\,The KK modes $B_{\mu}^{(n)}$ gain their mass by absorbing the degrees of freedom of the fictitious NG bosons $\varphi^{(n)}$ from the $w$-component of $A_M$.\,\,The NG boson equivalence theorem essentially states that the longitudinal component of the massive gauge field $B_{\mu}^{(n)}$ couples to the other fields in the same way as the fictitious NG boson $\varphi^{(n)}$ that it absorbed to gain mass.\,\,In abelian theories, this can be easily understood, as the two always appear in the combination $B_{\mu}^{(n)} - \frac{1}{m_n} \partial_{\mu} \varphi^{(n)}$.\,\,However, it is not difficult to show that the theorem also holds more generally in non-abelian theories.\,\,For readers interested in further details, we provide a brief review of the NG boson equivalence theorem in \hyperref[app:B_NG_Theorem]{Appendix B}.

Returning to the question of the scaling of scattering amplitudes involving $\rho$-mesons, notice that the wavefunction of the fictitious NG boson $\varphi^{(n)}$ is given by $\phi_{n} (w) \sim R_5^3/w^{3}$ in the UV-region, and its ``polarisation'' is $\zeta_M = (0, \ldots, 0, \zeta_w)$.\,\,Then, following the earlier power counting analysis from \hyperref[sec:4_1_Naive_Counting]{Section 4.1}, we find that scattering amplitudes with fictitious NG bosons $\varphi^{(n)}$ will scale in the same way as those with pions, and thus will not exhibit the suppression observed in scattering amplitudes with $\rho$-mesons.

Although the fictitious NG boson $\varphi^{(n)}$ is not a physical degree of freedom, the lack of suppression in scattering amplitudes involving these bosons still implies an inconsistency. The inconsistency arises because the NG boson equivalence theorem dictates that scattering amplitudes with longitudinally polarised $\rho$-mesons and fictitious NG bosons $\varphi^{(n)}$ should exhibit the same scaling with $s$ in the high-energy fixed-angle limit.\,\,However, as we found, a naive application of the original proposal \cite{Polchinski:2001tt} leads to a contradiction with this statement.

In the next subsection, we will show how our generalised proposal from \hyperref[sec:3_2_PS_General]{Section 3.2} recovers predictions consistent with the NG boson equivalence theorem and the constituent counting rule.
\subsection{Results from the Generalised Proposal} \label{sec:4_3_Scaling_Corrected}

Following our generalised proposal \eqref{eq:3_PS_Gen}, we argue that in the high-energy fixed-angle limit an $n$-meson scattering amplitude $\mathcal{A}_n$ to leading order in $s$ is given by:
\begin{align} \label{eq:A_n_meson_new_proposal}
    \mathcal{A}_{n}\Big(k_{\mu}^{(i)}, \zeta_{\mu}^{(i)}\Big) &= \int dw \sqrt{-g} \times \widetilde{\mathcal{S}}_{n}\Big(p_M^{(i)}, \zeta_M^{(i)}\Big) \times \prod_{i=1}^{n} \psi^{(i)}(w) \nonumber \\
    &= \int dw \sqrt{-g} \, \widetilde{F}_n^{M_1 N_1 \cdots M_n N_n} \Big(p_{\mu}^{(i)} \sqrt{\alp}\Big) \, \prod_{i=1}^{n} \widetilde{M}_{M_i N_i}^{(i)} \psi^{(i)}(w),
\end{align}
where the prescription given above holds for any $n$ only for U(1) mesons, while for $n=4$, it also applies to U($N_f$) mesons.\,\,For the definitions of $\widetilde{F}_n^{\cdots}$ and $\widetilde{M}_{M_i N_i}^{(i)}$, consult \hyperref[sec:3_2_PS_General]{Section 3.2}.

Here, recall that in $\widetilde{F}_n^{\cdots}$ and $\widetilde{M}_{M_i N_i}^{(i)}$, the four-dimensional momenta $p_{\mu}^{(i)}$ are replaced with $k_{\mu}^{(i)}$.\,\,Furthermore, in $\widetilde{F}_n^{\cdots}$, the components $p_w^{(i)}$ are set to zero for all $i$, while in $\widetilde{M}_{MN}^{(i)}$, the components $p_w^{(i)}$ are replaced with partial derivatives $-i \partial_w^{(i)}$, with each derivative acting exclusively on its corresponding wavefunction $\psi^{(i)}(w)$:
\begin{align}
    \widetilde{M}_{\mu \nu}^{(i)} \psi^{(i)}(w) &=  \Big(k_{\mu}^{(i)} \zeta_{\nu}^{(i)} - k_{\nu}^{(i)}\zeta_{\mu}^{(i)}\Big) \psi^{(i)}(w), \\
    \widetilde{M}_{\mu w}^{(i)} \psi^{(i)}(w) &= \Big(k_{\mu}^{(i)} \zeta_{w}^{(i)} + i \zeta_{\mu}^{(i)} \partial_w^{(i)}\Big) \psi^{(i)}(w).
\end{align}

Now, as discussed in \hyperref[sec:3_2_PS_General]{Section 3.2}, in the high-energy fixed-angle limit, the integral in equation \eqref{eq:A_n_meson_new_proposal} is dominated by the region where $\alp \tilde{s} \sim \mathcal{O}(1)$, as the superstring scattering amplitude exponentially suppresses contributions outside this region.\,\,In this region, $\widetilde{\mathcal S}_n$ is \emph{effectively} of order $\mathcal{O}(1)$ after acting on the wavefunctions $\psi^{(i)}(w)$.\,\,Thus, the leading scaling with $s$ of $\mathcal{A}_n$ can be found from the volume element and wavefunctions, as in~\hyperref[sec:4_1_Naive_Counting]{Section~4.1}. Moreover, with our generalised proposal, the scattering amplitudes with $\rho$-mesons no longer vanish because of additional contributions from the components $p_w^{(i)}$.\,\,Let us now precisely show why these scattering amplitudes do not vanish.\,\,To this end, we compute the action of $\widetilde{M}_{MN}^{(i)}$ on the wavefunctions $\psi^{(i)}(w)$, as presented in equation \eqref{eq:A_n_meson_new_proposal}.

For pions, where $\zeta_{\mu}^{(i)} = (0, \ldots, 0)$ and $\zeta_w^{(i)} \psi^{(i)}(w) \rightarrow \psi_{\pi}(w)$, the only contribution is:
\begin{equation} \label{eq:4_PionUV}
    \widetilde{M}_{\mu w}^{(i)} \psi^{(i)}(w) = k_{\mu}^{(i)} \psi_{\pi}(w) \sim k_{\mu}^{(i)} \, \frac{R_5^3}{|w|^3},
\end{equation}
where in the final expression above we focused on the UV-region and recall that the pion wavefunction $\psi_{\pi}(w)$ was given in equation \eqref{eq:wavefunction_pi}.

For $\rho$-mesons, the leading contribution will come from the longitudinal polarisations. Taking $\zeta_\mu^{(i)} = k_\mu^{(i)}/m_\rho$, we observe that $M_{\mu \nu}^{(i)}$ vanishes, as a consequence of gauge invariance. However, in our generalised proposal, $\zeta_M^{(i)}$ is no longer longitudinal in a five-dimensional sense, since at high energies $\zeta_M^{(i)} \approx \big(k_\mu^{(i)}/m_\rho, 0 \big) \neq p_M^{(i)}/m_{\rho}$, as $p_w^{(i)}$ is also included.

Therefore, in contrast to the result obtained through a naive application of the original proposal in \hyperlink{sec:4_2_PS_Vectors}{Section 4.2}, there is now an additional non-vanishing contribution given by:
\begin{equation} \label{eq:4_RhoUV}
    \widetilde{M}_{\mu w}^{(i)} \psi^{(i)}(w) = i \zeta_{\mu}^{(i)} \partial_w \psi_{\rho}(w) \sim \frac{k_\mu^{(i)}}{m_\rho} \frac{R_5^2}{w^3}.
\end{equation}
In the final expression above, we focused on the regime that provides the relevant result in the high-energy fixed-angle limit, specifically where $\zeta_{\mu}^{(i)} \approx k_\mu^{(i)}/m_\rho$ and $\psi_{\rho}(w) \sim R_5^2/w^2$.

Given that the scaling with $w$ and $s$ in equations \eqref{eq:4_PionUV} and \eqref{eq:4_RhoUV} matches, and there are no additional polarisation-dependent quantities in equation \eqref{eq:A_n_meson_new_proposal} beyond those already computed above, we conclude that the scattering amplitudes of pions and $\rho$-mesons will exhibit the same scaling with $s$ in the high-energy fixed-angle limit.\,\,In addition, as already mentioned above, by following the same power-counting analysis as in \hyperref[sec:4_1_Naive_Counting]{Section 4.1}, we can now faithfully reproduce the constituent counting rule \eqref{eq:4_Pred2}.

Finally, let us comment on the NG boson equivalence theorem.\,\,From our discussion, we find that the longitudinal component of the $\rho$-meson couples through the term:
\begin{equation} \label{eq:4_coupling1}
    \widetilde{M}_{\mu w}^{(i)} \psi^{(i)}(w) = i\frac{k_\mu^{(i)}}{m_\rho} \, \pa_w \psi_\rho(w).
\end{equation}
For comparison, consider the fictitious NG boson $\varphi^{(1)}$.\,\,Its polarisation is in the $w$-direction, similar to the pion, and the only coupling it has is:
\begin{equation} \label{eq:4_coupling2}
    \widetilde{M}_{\mu w}^{(i)} \psi^{(i)}(w) = i \frac{k_\mu^{(i)}}{m_\rho} \, \pa_w \psi_\rho(w),
\end{equation}
where we used the fact that the wavefunction of $\varphi^{(1)}$ is $\phi_1(w) = \frac{1}{m_{\rho}} \partial_w \psi_\rho(w)$, as written in equation \eqref{eq:2_Wave_Eq}.\,\,In conclusion, we observe that the couplings \eqref{eq:4_coupling1} and \eqref{eq:4_coupling2} are identical, reflecting gauge symmetry and indicating that the NG boson equivalence theorem is upheld.
\section{Summary and Outlook} \label{sec:5_Conclusions}

In this paper, we revisited the old problem of computing meson scattering amplitudes in the high-energy fixed-angle regime using the modern tools of string theory and holography. To this end, we generalised the Polchinski--Strassler proposal \cite{Polchinski:2001tt}, originally formulated for glueballs, and applied it to investigate meson scattering in a similar spirit to reference \cite{Bianchi:2021sug}. Our focus was on the scattering amplitudes of pions and $\rho$-mesons.\,\,However, our findings can also be extended to the scattering of the other vector and axial vector mesons captured by the holographic QCD model from \hyperref[sec:2_HQCD_Model]{Section 2}.

We pointed out that a naive application of the original Polchinski--Strassler proposal~\cite{Polchinski:2001tt} to study the scattering amplitudes with $\rho$-mesons does not lead to the expected constituent counting rule.\,\,Nevertheless, the original proposal was sufficient to reproduce the constituent counting rule for scattering amplitudes with pions.\,\,The failure to reproduce the expected behaviour for scattering amplitudes with $\rho$-mesons was due to the scattering amplitude vanishing when the $\rho$-meson polarisation was chosen to be proportional to its momentum. This property followed directly from gauge invariance of the open superstring scattering amplitudes and hence was unavoidable in our holographic model, in which the $\rho$-meson is realised as a component of the massless five-dimensional gauge field $A_M$.

More importantly, the scaling behaviour of the $\rho$-meson scattering amplitudes, obtained from a naive application of the original proposal \cite{Polchinski:2001tt}, was inconsistent with the NG boson equivalence theorem, which asserts the equivalence between amplitudes involving longitudinal $\rho$-mesons and those involving fictitious NG bosons in the high-energy limit. Applying the original proposal in the context of our holographic model showed that scattering amplitudes with fictitious NG bosons scale with $s$ in the same way as those involving pions, since both the fictitious NG boson and the pion arise from the same component $A_w$ of $A_M$, and their wavefunctions scale identically with $w$ in the UV-region.\,\,This resulted in an inconsistency with the NG boson equivalence theorem, as naively applying the original proposal led us to predict that scattering amplitudes involving pions did not scale with $s$ in the same way as those involving $\rho$-mesons.

The main point of our proposal was to retain the $w$-derivative $\partial_w^{(i)}$ in the factor $\widetilde{M}_{MN}^{(i)}$, e.g.\,\,written in equation \eqref{MMN}.\,\,Moreover, we argued that the terms involving $\partial_w^{(i)}$ in $\widetilde{M}_{MN}^{(i)}$ contribute the leading term in the limit $s \to \infty$ for $\rho$-meson scattering amplitudes and thus cannot be neglected.\,\,With this approach, both the expected constituent counting rule for $\rho$-meson scattering amplitudes and the NG boson equivalence theorem were recovered.

To obtain the constituent counting rule from our generalised proposal for $n$-point scattering amplitudes, we focused on the case with $N_f = 1$.\,\,In this case, the dependence on polarisation in the scattering amplitude can be expressed in terms of the antisymmetric tensor $M_{MN}^{(i)} = p_M^{(i)} \zeta_N^{(i)} - p_N^{(i)} \zeta_M^{(i)}$, as shown in equation \eqref{eq:S_n_generic}.\,\,This factorisation also holds for the specific case of $n = 4$ with generic $N_f$.

In \hyperref[sec:3_2_PS_General]{Section 3.2}, this property was needed to argue that, assuming $\alp/R_5^2 \ll 1$, we can neglect all factors of $p_w^{(i)}$ present in the superstring scattering amplitude outside of~$M_{MN}^{(i)}$. The antisymmetry of $M_{MN}^{(i)}$ also enabled us to replace all covariant derivatives $\nabla^{(i)}_M$ in $M_{MN}^{(i)}$ with ordinary derivatives: $p_M^{(i)} \to \big(k_\mu^{(i)},-i\pa_w^{(i)}\big)$.\,\,Without this property, our proposal can, in principle, suffer from ordering ambiguities due to the non-vanishing of the commutator of the covariant derivatives $\big[\nabla_M^{(i)},\nabla_N^{(i)}\big] \neq 0$.\footnote{Here, recall that the commutator $\big[\nabla_M^{(i)}, \nabla_N^{(j)}\big]$ for $i \neq j$ vanishes because the derivatives act exclusively on their respective wavefunctions.}\,\,Given this, our proposal might not be fully defined in the most general case of $N_f > 1$ and any $n$ without resolving these ambiguities.

Expressing the scattering amplitudes in terms of $M_{MN}^{(i)}$ also allowed us to easily recover the NG boson equivalence theorem.\,\,However, for $N_f > 1$, the non-abelian structure may introduce additional terms.\,\,One counterexample is the 3-point function \cite{Green:1987sp}, and another is the 5-point function \cite{Medina:2002nk,Barreiro:2005hv}.\,\,These scattering amplitudes with odd $n$ vanish for $N_f = 1$, but have non-zero expressions for $N_f > 1$, which are not of the form of equation \eqref{eq:S_n_generic}.

To see where the difficulty might arise in the $n$-point scattering amplitude with $N_f > 1$, let us outline how agreement with the NG boson equivalence theorem might be recovered. In general, a flat-space scattering amplitude $\mathcal{S}_n$ of massless U($N_f$) gauge bosons is~given~by:
\begin{equation}
    {\cal S}_n = {\cal S}_n^{M_1 \cdots M_n} \zeta^{(1)}_{M_1} \cdots \zeta^{(n)}_{M_n}.
\end{equation}
In the previous equation, $\mathcal{S}_n^{\cdots}$ is a tensor constructed from the momenta and the Minkowski metric, and each polarisation $\zeta_M^{(i)}$ appears exactly once for $i = 1, 2, \ldots, n$.

Now, gauge invariance implies that ${\cal S}_n$ should vanish when one of the polarisations, without loss of generality $\zeta^{(1)}_M$, is taken to be proportional to its respective momentum~$p^{(1)}_M$. Then, by replacing the five-dimensional polarisation as $\zeta_M^{(1)} \rightarrow \frac{1}{m_\rho} \big(k_\mu^{(1)}, p_w^{(1)}\big)$, we arrive at:
\begin{equation} \label{eq:gauge_inv_general}
    \mathcal{S}_n^{\mu M_2 \cdots M_n} \frac{k_{\mu}^{(1)}}{m_\rho} \zeta_{M_2}^{(2)} \cdots \zeta_{M_n}^{(n)} = -\mathcal{S}_n^{w M_2 \cdots M_n} \frac{p_w^{(1)}}{m_\rho} \zeta_{M_2}^{(2)} \cdots \zeta_{M_n}^{(n)}.
\end{equation}
We identify the term on the left-hand side as the scattering amplitude with the longitudinal $\rho$-meson, where we set $\zeta_M^{(1)} = \frac{1}{m_{\rho}} \big(k_{\mu}^{(1)}, 0\big)$, while the term on the right-hand side is identified as the amplitude with the fictitious NG boson, with $\zeta_M^{(1)} = \big(0, \dots, 0, \zeta_w^{(1)}\big)$.\,\,In our proposal, the component $p_w^{(1)}$ will act on the wavefunction $\psi^{(1)}(w)$, to yield $\partial_w \psi_{\rho}(w) = m_{\rho} \phi_1(w)$. The statement of gauge symmetry is that the two scattering amplitudes above are equal, and from this, in principle, we should be able to recover the constituent counting rule for all scattering amplitudes, following the power counting analysis from \hyperref[sec:4_1_Naive_Counting]{Section 4.1}.

The issue with the generic, non-abelian $n$-point scattering amplitude is that, due to potential ambiguities in our proposal, it is not clear whether the scattering amplitudes on both sides of equation \eqref{eq:gauge_inv_general} are well-defined without additional input.

Despite these subtleties, since our findings hold for non-abelian 4-point functions and there is no reason to believe that gauge symmetry will cause scattering amplitudes to vanish for $n > 4$, it is not unreasonable to expect that our proposal will continue to reproduce the constituent counting rule for non-abelian $n$-point scattering amplitudes involving pions and $\rho$-mesons:
\begin{equation}
    \mathcal{A}_n \sim s^{2-n} f_n(\theta_i),
\end{equation}
where $f_n(\theta_i)$ can, in principle, be computed using our proposal, given the explicit form of the $n$-point open superstring scattering amplitude in Minkowski space.

Our approach can be extended to many other types of mesons and scattering processes. In particular, it is straightforward to compute scattering amplitudes that involve higher spin mesons by realising them as excited open string states \cite{Imoto:2010ef}.\,\,Here, it is also worth noting that our holographic QCD model already includes the $a_1$-meson, which is captured by the mode $B_{\mu}^{(2)}$.\,\,Therefore, it would be worthwhile to examine the scattering amplitudes involving these mesons in more detail using our proposal.

Although holographic models of QCD have shown promise in fitting a wide range of empirical data, computations of scattering processes remain less common in the literature. Existing works often focus on qualitative aspects, including proton-proton scattering via pomeron exchange and reggeization \cite{Hu:2017iix}, pion-pion scattering \cite{Bianchi:2020cfc, Veneziano:2017cks,Hoyos:2022ptd}, reggeization \cite{Domokos:2009hm}, holographic corrections \cite{Armoni:2016llq, Armoni:2016nzm, Armoni:2017dcr}, and pomeron/reggeon dynamics \cite{Polchinski:2002jw, Brower:2006ea, Amorim:2021gat}.\,\,Albeit, a key advantage of our approach is that it can be used to compute the full angular dependence of the studied meson scattering amplitudes.\,\,In contrast to the constituent counting rule, which is universal, the angular behaviour depends on the specifics of the holographic model.
\newpage \noindent As we have demonstrated, the constituent counting rule is recovered from the UV-region, where the bulk geometry is asymptotically AdS and the meson wavefunctions decay with the appropriate power-law behaviour.\,\,However, the expression for the angular dependence, given in equation \eqref{eq:4_A_n_pion_rho}, requires as input the entire bulk geometry and wavefunctions, including their behaviour in the IR-region.

Most progress made in computing scattering amplitudes and correlation functions in holography tends to rely heavily on supersymmetry and conformal symmetry, limiting their applicability to non-supersymmetric, confining geometries required for holographic~QCD. Of particular note is the tangible progress made in computing closed superstring scattering amplitudes in AdS space \cite{Alday:2023mvu}, where the high-energy fixed-angle \cite{Alday:2023pzu} and Regge limits~\cite{Alday:2024xpq} have been investigated.\,\,In addition, open superstring scattering amplitudes in AdS space were also explored in reference \cite{Alday:2024yax}.\,\,It will be interesting to explore if our generalisation of the Polchinski--Strassler proposal \cite{Polchinski:2001tt} can be connected to these works.

A great challenge for string theory is to describe baryon scattering.\,\,Hadron colliders, such as the LHC, scatter protons, not pions.\,\,It is an important and challenging task to describe such collisions, because in string theory baryons are described by D-branes, which are non-perturbative objects.\,\,We postpone this very interesting problem to future studies (see references \cite{Domokos:2009hm,Hu:2017iix} for works in this direction).
\acknowledgments{

We thank S. J. Brodsky and H. Kawamura for useful correspondence.\,\,BP also thanks Z.~Du, Y. Fu, K. Ikeda, S. Moriyama, M. Ward, and X. U. Nguyen for insightful discussions.

AA was supported by STFC grant ST/T000813/1.\,\,BP was supported by JST SPRING, Grant Number JPMJSP2110.\,\,The work of SS was supported by the JSPS KAKENHI (Grant-in-Aid for
Scientific Research (B)) grant number JP24K00628 and MEXT KAKENHI (Grant-in-Aid
for Transformative Research Areas A ``Extreme Universe'') grant number 21H05187.\,\,DW was supported by an appointment to the YST Program at the APCTP through the Science and Technology Promotion Fund and Lottery Fund of the Korean Government.\,\,This was also supported by the Korean Local Governments of Gyeongsangbuk-do Province and Pohang City.}
\clearpage
\appendix
\section{Conventional Derivation of the Constituent Counting Rule} \label{app:A_Scaling_Rules}

The constituent counting rule \eqref{eq:1_Scaling_Rule} can be derived from conventional dimensional analysis. Here, following the original references \cite{Matveev:1973ra, Brodsky:1973kr} and the more recent work \cite{Kawamura:2013iia}, we will briefly review this derivation.\,\,The constituent counting rule discussed here is expected to hold in asymptotically free confining gauge theories in four dimensions.\footnote{In confining theories with operators that have non-vanishing anomalous dimensions in the UV-limit, we expect the exponent of $s$ in equation \eqref{eq:A_Scaling} to receive additional corrections (e.g.\,\,see reference \cite{Callan:1974zy}).}\,\,Our approach below follows an argument from perturbative QCD in four dimensions, and treats the asymptotic hadron states as free quarks moving together.\footnote{More details on how non-perturbative effects can be incorporated are provided in reference \cite{Brodsky:1974vy}.}

\begin{figure}[hb!] \label{fig:A_Quark_Diagram}
    \centering
    \includegraphics[scale = 1]{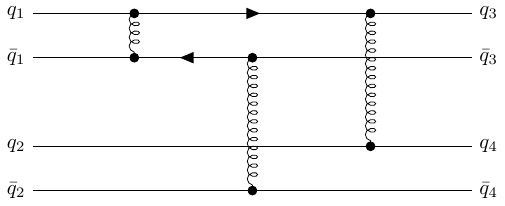}
    \caption{A minimal Feynman diagram for 2-to-2 meson scattering, where $q_i$ and ${\bar q}_i$ represent quarks and antiquarks for $i = 1, \ldots, 4$.\,\,In agreement with the counting above, this diagram has $m = 8$ external lines, 3 gluon propagators, 2 internal fermion propagators, and 6 interaction vertices.}
\end{figure}

In the high-energy fixed-angle regime, there is only a single relevant energy scale given by the momentum $P \sim \sqrt{s}$. For hadron scattering involving $m$ constituent (anti)quarks, we can focus solely on diagrams where each quark line is connected by at least one gluon, using the fewest gluons necessary.\,\,These diagrams will scale with the highest power of $s$ in the high-energy fixed-angle regime.\,\,An example of such a diagram is shown in \hyperref[fig:A_Quark_Diagram]{Figure 2}.

In general, a minimal diagram includes:
\begin{itemize}
    \item $\frac{m}{2}-1$ gluon propagators of dimension $P^{-2}$,
    \item $\frac{m}{2}-2$ (anti)quark propagators of dimension $P^{-1}$,
    \item $m$ external fermions whose normalised wavefunctions carry dimension $\sqrt{P}$,
    \item $m-2$ interaction vertices with a dimensionless coupling constant.
\end{itemize}
Multiplying these factors together, we find that the amplitude will scale as $P$ to the power:
\begin{equation}
    -2 \times \bigg(\frac{m}{2}-1\bigg) - 1 \times \bigg(\frac{m}{2}-2\bigg) + \frac{1}{2} \times m = 4-m.
\end{equation}
Thus, we find that a scattering amplitude of $n = m/2$ mesons must scale as:
\begin{equation} \label{eq:A_Scaling}
    {\cal A} \sim s^{2-\frac{m}{2}} \sim s^{2-n}.
\end{equation}

\section{The Nambu--Goldstone Boson Equivalence Theorem} \label{app:B_NG_Theorem}

In this appendix, we briefly review the NG boson equivalence theorem, mainly following Chapter 21.2 of reference \cite{Peskin:1995ev}.\,\,For the complete original proof, see references \cite{Cornwall:1974km, Vayonakis:1976vz}.

In gauge theories with spontaneously broken gauge symmetries, the physical behaviour of massive vector bosons remains closely controlled by the original gauge symmetry, even after symmetry breaking.\,\,To illustrate this, we will examine an arbitrary matrix element that involves a gauge current between two on-shell states $\mathcal{M}_{\mu} = \bra{\text{on-shell}} J_{\mu}(k_{\mu}) \ket{\text{on-shell}}$, where $k_{\mu}$ is the external momentum of the gauge current $J_{\mu}$.\,\,We choose the Lorentz gauge (i.e.\,\,the $R_{\xi}$ gauge with $\xi = 0$), which includes the fictitious NG boson that gives mass to the external vector boson (related to the gauge current $J_{\mu}$) after symmetry breaking.

In the Lorentz gauge, the Ward identity reads as follows:
\begin{equation} \label{eq:A_Generic_M}
    0 = k_{\mu} \mathcal{M}^{\mu} = k_{\mu} \mathcal{M}_1^{\mu} + k_{\mu} F^{\mu} \mathcal{M}_2,
\end{equation}
where $\mathcal{M}_1^{\mu}$ denotes a one-particle irreducible (1PI) vertex and $\zeta_{\mu} \mathcal{M}_1^{\mu}$ denotes a scattering amplitude that includes the external vector boson with momentum $k^{\mu}$ and polarisation $\zeta^{\mu}$. Furthermore, $F^{\mu} \mathcal{M}_2$ represents a contribution from a diagram in which the propagator of the fictitious NG boson is connected to the external vector boson leg.\,\,Here, $\mathcal{M}_2$ denotes the scattering amplitude with the fictitious NG boson, with the external vector boson leg stripped off, and $F^{\mu}$ contains the propagator of the fictitious NG boson.

In addition, notice that $F^{\mu}$ can be determined by appealing to Lorentz covariance and recognising that it contributes to the mass $m_A$ in the propagator of the external vector boson after symmetry breaking, giving $F^{\mu} = - m_A k^{\mu} / k^2$.

In the high-energy limit where the momentum of the external vector boson $|\vec{k}| \to \infty$, the longitudinal polarisation $\zeta_{\mu}^L$ of the vector boson becomes equal to $k_{\mu} / m_A$ with subleading corrections of order $\mathcal{O}(m_A/k_0)$.\,\,Thus, by taking the high-energy limit of equation~\eqref{eq:A_Generic_M} and using the identity $F^{\mu} = - m_A k^{\mu} / k^2$, we obtain:
\begin{equation}
    \zeta_{\mu}^L \mathcal{M}_1^{\mu} = \mathcal{M}_2 \times \big(1 + \mathcal{O}\big(m_A^2/k_0^2\big) \big).
\end{equation}
The equation above is known as the NG boson equivalence theorem.

\begin{figure}[hb!] \label{fig:Equivalence_Diagram}
    \centering
    \includegraphics[scale = 1]{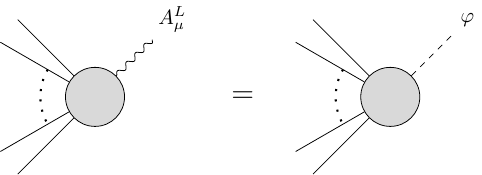}
    \caption{The NG boson equivalence theorem, at leading order in the high-energy limit, relates the amplitudes for absorbing or emitting a longitudinally polarised vector boson $A_{\mu}^L$ to those for absorbing or emitting a corresponding NG boson $\varphi$ eaten by $A_{\mu}^L$ after symmetry breaking.}
\end{figure}
\section{Open Superstring Scattering Amplitudes in Minkowski Space} \label{app:C_pw_zero}

In this Appendix, we explicitly demonstrate that $F_n^{\mu_1 w \cdots \mu_n w}$ is non-zero, a property that is relevant at the end of \hyperref[sec:3_2_PS_General]{Section 3.2}, just below equation \eqref{eq:3_SecondT}.\,\,In our analysis, we follow the conventions from the review in reference \cite{Medina:2002nk}.

The tree-level $n$-point open superstring scattering amplitude of U($N_f$) gauge bosons in ten-dimensional Minkowski space is given by \cite{Green:1987sp,Polchinski:1998rr, Medina:2002nk}:
\begin{equation}
    \mathcal{S}_{n} \Big(p^{(i)}, \zeta^{(i)}\Big) = i (2 \pi)^{10} \, \delta^{(10)} \Big({\textstyle \sum_{i=1}^n p^{(i)}}\Big) \sum_{\rm perm}{\rm tr}(\lambda^{a_1} \lambda^{a_2} \cdots \lambda^{a_n}) \, A(1, 2, \dots, n),
\end{equation}
where $\textstyle \sum_{\rm perm}$ denotes the sum over all non-cyclic permutations of the sets $\{p^{(i)}, \zeta^{(i)}, a_i\}$ for $i = 1, 2, \dots, n$, and $\lambda^{a_i}$ are the U($N_f$) generators associated with the $i$-th gauge boson. Moreover, $A(1, 2, \dots, n)$ is expressed by:
\begin{align} \label{An}
    A(1, 2, \dots, n) = \, \, & 2 g_s^{n-2} \, (x_{n-1}-x_1)(x_n-x_1) \times \nonumber \\
    &\times \int dx_2 \cdots dx_{n-2} \int d\theta_1 \cdots d\theta_{n-2} \, \prod_{i>j}^n (x_i-x_j-\theta_i\theta_j)^{p^{(i)} \cdot p^{(j)}} \times \nonumber \\
    &\times \int d\phi_1 \cdots d\phi_n \exp\Bigg(\sum_{i \ne j}^n \frac{(\theta_i-\theta_j) \phi_i \, \zeta^{(i)} \cdot p^{(j)}-\frac12 \phi_i \phi_j \, \zeta^{(i)} \cdot \zeta^{(j)}}{x_i-x_j-\theta_i\theta_j}\Bigg),
\end{align}
where $x_i$ are real parameters that satisfy $x_1 < x_2 < \cdots < x_n$, whereas $\theta_i$ and $\phi_i$ are Grassmannian variables, and we have set $\alp = 1/2$.\,\,Notice that the superstring scattering amplitude above does not depend on $x_1$, $x_{n-1}$, $x_n$, $\theta_{n-1}$, and $\theta_n$.\,\,Thus, to simplify our discussion, we set $(x_1, x_{n-1}, x_n) = (0, 1, \infty)$ and $\theta_{n-1} = \theta_n = 0$.

Let us now examine the case where the only non-zero components of the momenta $p^{(i)}$ and polarisations $\zeta^{(i)}$ for all $i$ are $p_\mu^{(i)}$ for $\mu = 0, \dots, 3$ and $\zeta_w^{(i)} := \zeta_4^{(i)} = 1$, respectively. In~this case, the non-zero components of $M_{MN}^{(i)} = p_M^{(i)} \zeta_N^{(i)} - p_N^{(i)} \zeta_M^{(i)}$ are $M_{\mu w}^{(i)}=-M_{w\mu}^{(i)}=p_\mu^{(i)}$, and according to equation \eqref{eq:S_n_generic}, the scattering amplitude $\mathcal{S}_{n}$ for $N_f=1$ in this setup is:
\begin{align} \label{SnC}
        \mathcal{S}_n \Big(p^{(i)}, \zeta^{(i)}\Big) = 2^n F_n^{\mu_1 w \cdots \mu_n w} \Big(p_\mu^{(i)} \sqrt{\alp}\Big) \, M_{\mu_1 w}^{(1)} \cdots M_{\mu_n w}^{(n)}.
\end{align}
Then, if we could establish that the scattering amplitude above is non-vanishing, we would conclude that $F_n^{\mu_1 w \cdots \mu_n w}$ is non-zero when the components $p^{(i)}_w$ are set to zero for all $i$.

To show that the amplitude \eqref{SnC} is non-vanishing, we evaluate the expression \eqref{An} more explicitly.\,\,With our chosen momenta and polarisations, we find that $\zeta^{(i)} \cdot p^{(j)} = 0$ and $\zeta^{(i)} \cdot \zeta^{(j)} = 1$.\,\,As a result, the integrand in equation \eqref{An} simplifies to:
\begin{align} \label{int}
    &\prod_{i>j}^{n}(x_i-x_j-\theta_i\theta_j)^{p^{(i)} \cdot p^{(j)}} \exp\bigg(-\frac{\phi_i\phi_j}{x_i-x_j-\theta_i\theta_j}\bigg) \nonumber \\
    = &\prod_{i>j}^{n}(x_i-x_j)^{p^{(i)} \cdot p^{(j)}} \exp\bigg(-\frac{p^{(i)} \cdot p^{(j)} \, \theta_i\theta_j + \phi_i\phi_j}{x_i-x_j}-\frac{\phi_i\phi_j \, \theta_i\theta_j}{(x_i-x_j)^2}\bigg) \nonumber \\
    \sim &-\frac{1}{x_n} \phi_n\phi_{n-1} \prod_{i>j}^{n-1}(x_i-x_j)^{p^{(i)} \cdot p^{(j)}} \exp \Big( -A_{ij}\theta_i\theta_j-B_{ij}\phi_i\phi_j-C_{ij}\phi_i\phi_j\theta_i\theta_j \Big).
\end{align}
In the previous equation:
\begin{align}
    A_{ij} = \frac{p^{(i)} \cdot p^{(j)}}{x_i-x_j}, \qquad B_{ij} = \frac{1}{x_i-x_j} - \frac{1}{1-x_j} + \frac{1}{1-x_i}, \qquad C_{ij} = \frac{1}{(x_i-x_j)^2}.
\end{align}
In the final step of equation \eqref{int}, we extracted the term proportional to $\phi_n \phi_{n-1}$ and used the identity $p^{(n)} \cdot \big(p^{(1)} + \cdots + p^{(n-1)}\big) = 0$, which follows from momentum conservation and the mass-shell condition $p^{(i)} \cdot p^{(i)} = 0$.\,\,Notice that the factor $1/x_n$ in the expression \eqref{int} is cancelled by the factor $(x_{n-1}-x_1)(x_n-x_1) = x_n$ in equation \eqref{An}.

Substituting the expression \eqref{int} into $A(1, 2, \dots, n)$ in equation \eqref{An} and performing the fermionic integrals, we arrive at:
\begin{align} \label{An2}
    &A(1, 2, \dots, n) = \frac{g_s^{n-2}}{2^{n-3}} \int dx_2 \cdots dx_{n-2} \prod_{i>j}^{n-1}(x_i-x_j)^{p^{(i)} \cdot p^{(j)}} \sum_{\substack{j_1,\dots,j_{n-2}=1\\{\rm distinct}}}^{n-2}\sum_{k=0}^{\frac{n}{2}-1}\frac{2^{\frac{n}{2}-1-k}}{(k!)^2\left(\frac{n}{2}-1-k\right)!} \nonumber \\
    &\times \sum_{\sigma\in S_{2k}}{\rm sgn}(\sigma) A_{j_1j_2}\cdots A_{j_{2k-1}j_{2k}}B_{j_{\sigma(1)}j_{\sigma(2)}}\cdots B_{j_{\sigma(2k-1)}j_{\sigma(2k)}}C_{j_{2k+1}j_{2k+2}}\cdots C_{j_{n-3}j_{n-2}}.
\end{align}
In the equation above, $S_{2k}$ denotes the symmetric group of order $2k$, ${\rm sgn}(\sigma)$ is the sign of the permutation $\sigma \in S_{2k}$, integration over $x_i$ is restricted to $0 < x_2 < \cdots < x_{n-2} < 1$, and summation over $j_i$ for $i = 1, \dots, n-2$ is taken over distinct values.

Examining equation \eqref{An2}, it is not difficult to see that it is non-zero for a generic choice of momenta.\,\,This implies that whenever $\mathcal{S}_n$ admits a factorisation of the form given in equation \eqref{eq:S_n_generic}, as is the case for $n$-point scattering of U(1) gauge bosons and 4-point scattering of U($N_f$) gauge bosons (omitting Chan--Paton factors), our reasoning shows that $F_n^{\mu_1 w \cdots \mu_n w}$ is non-zero for a generic choice of momenta, thereby completing our argument. In summary, recall that $F_n^{\mu_1 w \cdots \mu_n w}$ is independent of polarisations, which makes this result relevant to both scattering processes involving pions and $\rho$-mesons in our proposal.
\section{Longitudinal Polarisations and Scaling with Energy} \label{app:D_zeta_choice}

Consider the rest frame of a massive vector (e.g.\,\,a $\rho$-meson) with momentum:
\begin{equation}
    k_{\mu}^{\text{r.f.}} = (m, 0, 0, 0),
\end{equation}
where $m$ is the mass of the massive vector.\,\,For any such massive vector, the polarisation vector in its rest frame can be written as follows:
\begin{equation}
    \zeta_{\mu}^{\text{r.f.}} = (0, \sin{\phi}, \cos{\phi} \cos{\lambda}, \cos{\phi} \sin{\lambda}),
\end{equation}
which satisfies $k^{\text{r.f.}} \cdot \zeta^{\text{r.f.}} = 0$ and $\zeta^{\text{r.f.}} \cdot \zeta^{\text{r.f.}} = 1$.\,\,The polarisation vector has three independent polarisation modes parametrised by the choice of two angles $\phi$ and $\lambda$.\,\,In the rest frame, the polarisation is always of $\mathcal{O}(1)$.

Now, we assume that in the lab frame the massive vector is moving in the $x^1$-direction with energy $E$, such that its momentum is given by:
\begin{equation}
    k_{\mu}^{\text{l.f.}} = \bigg(E,\sqrt{E^2-m^2}, 0, 0\bigg).
\end{equation}
To reach this frame, we have boosted $k_{\mu}^{\text{r.f.}}$ in the $x^1$-direction with rapidity $w = \cosh^{-1}\frac{E}{m}$. After applying the same transformation to $\zeta_{\mu}^{\text{r.f.}}$, we obtain the polarisation in the lab frame:
\begin{equation}
    \zeta_{\mu}^{\text{l.f.}} = \bigg(\frac{\sqrt{E^2-m^2}}{m} \sin{\phi}, \frac{E}{m} \sin{\phi}, \cos{\phi} \cos{\lambda}, \cos{\phi} \sin{\lambda}\bigg).
\end{equation}
From the expression above, we can verify that while $\zeta_{\mu}^{\text{l.f.}}$ remains transverse in the sense that $\zeta^{\text{l.f.}} \cdot k^{\text{l.f.}} = 0$, in the high-energy limit, it becomes parallel to the momentum $k_{\mu}^{\text{l.f.}}$, as~$\zeta_{\mu}^{\text{l.f.}} \to \frac{\sin{\phi}}{m} k_{\mu}^{\text{l.f.}}$, and is of order $E \sim \sqrt{s}$ in the context of high-energy fixed-angle scattering processes.\,\,The only situation in which this does not occur is when we begin with a purely transverse polarisation with $\phi = 0$, orthogonal to the direction of the momentum, although that is not the case in general.

\bibliographystyle{JHEP}
\bibliography{biblio}

\end{document}